\documentclass[useAMS,usenatbib]{mnras} 
\usepackage{color}
\usepackage{times,psfig,epsfig} 
\usepackage{rotating, graphicx, adjustbox}
\usepackage{amssymb, amsmath}
\usepackage[T1]{fontenc} 
\usepackage{aecompl}
\usepackage{gensymb}
\usepackage{bm}

\usepackage{comment}

\newcommand{\vb}[1]{\mathbf{#1}}
\newcommand{\dv}[2]{\frac{\mathrm{d}#1}{\mathrm{d}#2}}
\newcommand{\pdv}[2]{\frac{\partial #1}{\partial #2}}
\newcommand{\dd}{\mathrm{d}}

\title[Turbulence from the ICM assembly history]{Troubled cosmic flows: turbulence, enstrophy and helicity from the assembly history of the intracluster medium}
\author[Vall\'es-P\'erez, Planelles \& Quilis]
 {David Vall\'es-P\'erez$^{1}$\thanks{david.valles-perez@uv.es}, Susana Planelles$^{1,2}$ and Vicent Quilis$^{1,2}$\\ 
  $^1$Departament d'Astronomia i Astrof\'{\i}sica, Universitat de
  Val\`encia, E-46100 Burjassot (Val\`encia), Spain\\ 
  $^{2}$Observatori Astron\`omic, Universitat de Val\`encia, E-46980
  Paterna (Val\`encia), Spain} 
\date{Accepted 2021 March 24. Received 2021 March 23; in original form 2021 February 19}

\pagerange{\pageref{firstpage}--\pageref{lastpage}} \pubyear{2021}

\def\LaTeX{L\kern-.36em\raise.3ex\hbox{a}\kern-.15em
    T\kern-.1667em\lower.7ex\hbox{E}\kern-.125emX}

\binoppenalty=\maxdimen
\relpenalty=\maxdimen

\begin{document}

\label{firstpage}
\maketitle


\begin{abstract}
Both simulations and observations have shown that turbulence is a pervasive phenomenon in cosmic scenarios, yet it is particularly difficult to model numerically due to its intrinsically multiscale character which demands high resolutions. Additionally, turbulence is tightly connected to the dynamical state and the formation history of galaxies and galaxy clusters, producing a diverse phenomenology which requires large samples of such structures to attain robust conclusions. In this work, we use an adaptive mesh refinement (AMR) cosmological simulation to explore the generation and dissipation of turbulence in galaxy clusters, in connection to its assembly history. We find that major mergers, and more generally accretion of gas, is the main process driving turbulence in the intracluster medium. We have especially focused on solenoidal turbulence, which can be quantified through enstrophy. Our results seem to confirm a scenario for its generation which involves baroclinicity and compression at the external (accretion) and internal (merger) shocks, followed by vortex stretching downstream of them. We have also looked at the infall of mass to the cluster beyond its virial boundary, finding that gas follows trajectories with some degree of helicity, as it has already developed some vorticity in the external shocks.
\end{abstract}

\begin{keywords}
hydrodynamics – turbulence – galaxies: clusters: general – galaxies: clusters: intracluster medium – methods: numerical
\end{keywords}


\section{Introduction}
\label{s:intro}

Within the hierarchical paradigm of cosmological structure formation (\citealp{PressSchechter_1974, GottRees_1975}; see also \citealp{KravtsovReview_2012} and \citealp{PlanellesReview_2015} for recent reviews), the assembly history of galaxy clusters is dominated by (major) merger events, which account for most of their mass growth (e.g., \citealp{Valles_2020}) and have consequential effects on their thermal structure (e.g., \citealp{Planelles_2009, ZuHone_2011}). Galaxy cluster mergers and, more generally, gas accretion are the main energy source fuelling different complex hydrodynamical processes in the intracluster medium (ICM), such as turbulence and a rich phenomenology of shock waves (see, e.g., \citealp{Quilis_1998, Ryu_2003, Zhang_2020}). Thus, a precise understanding of these phenomena is necessary for a proper description of the bulk of clusters' baryonic mass, which, furthermore, is essential for the usage of galaxy clusters as cosmological probes \citep[for a review]{Allen_2011} and to correctly model galaxy formation processes in a cosmological context \citep{Naab_2017}.

Constraining the amplitude and spectrum of turbulence, and elucidating their evolution with cosmic time and possible dependencies on clusters' mass, dynamical state, formation history, etc. is a fundamental step towards a correct description of many phenomena: non-thermal pressure \citep{Nelson_2014, Shi_2014, Vazza_2018, Angelinelli_2020} leading to hydrostatic mass bias \citep{Nelson_2014_massbias, Biffi_2016, Shi_2016}, amplification of cosmic magnetic fields (\citealp{Subramanian_2006, Iapichino_2008, Cho_2014, Beresnyak_2016, Vazza_2018_b, Brzycki_2019}; see also \citealp{Donnert_2018} for a recent review), chemical and thermal mixing \citep{Ruszkowski_2010, ZuHone_2011, Shi_2020}, star formation \citep{Kretschmer_2020}, cosmic-ray acceleration and radio emission \citep{Fujita_2003, Cassano_2005, Brunetti_2011, Brunetti_2020}, and heating due to viscous dissipation, which could help to alleviate the cooling flow problem \citep{Zhuravleva_2014_nature, Valdarnini_2019, Shi_2020}. 

Turbulence can be sourced, not only from gas accretion and mergers, but also due to galaxy motions \citep{Faltenbacher_2005, Kim_2007, Ruszkowski_2011}, feedback from active galactic nuclei (AGNs; \citealp{Gaspari_2012, Gaspari_2018}; cf. \citealp{Sayers_2021}), interaction of a compact, cool core with the surrounding ICM \citep{Valdarnini_2011, Valdarnini_2019}, etc. According to the classical \cite{Kolmogorov_1941} model for fully developed, homogeneous and isotropic turbulence, bulk motions induced by these large-scale structure (LSS) formation processes break down to smaller scales due to different instabilities, transferring energy from the injection scale (the characteristic scale of bulk motions) down to the dissipation scale, where this energy is converted to heat, magnetic field amplification and cosmic ray acceleration, among others. While this model could be rather idealized (the ICM is far from homogeneous; see \citealp{Shi_2018, Shi_2019, Mohapatra_2020, Mohapatra_2021} for several recent exploratory studies on stratified, ICM-like turbulence), it constitutes a simple baseline for exploring the turbulent phenomena in galaxy clusters.

Direct observations of turbulent flows in galaxy clusters have yet been elusive, in part due to the unfortunate fate of the Hitomi mission \citep{Kitayama_2014}. Indirect detections of turbulence have been reported by measurements of X-ray surface brightness fluctuations (e.g., \citealp{Churazov_2012, Zhuravleva_2014_nature}; see also \citealp{Gaspari_2013, Zhuravleva_2014_ApJL} for theoretical investigations) or Sunyaev-Zel'dovich (SZ) signal fluctuations \citep{Khatri_2016}. We refer the interested reader to \cite{Simionescu_2019} for a recent review on the current constraints and detectability prospects of gas motions in the ICM. In the next one to two decades, future X-ray missions (\textsc{Xrism}\footnote{\url{http://xrism.isas.jaxa.jp/en/}}, \textsc{Athena}\footnote{\url{https://www.the-athena-x-ray-observatory.eu/}}) will offer an unprecedented level of insight onto the ICM turbulent motions. Potentially, future mm/sub-mm facilities (e.g., \textsc{AtLAST}, see \citealp{Pamela_2020}; or \textsc{SKA}\footnote{\url{https://www.skatelescope.org/}}, \citealp{AcostaPulido_2015}) will also be able to map the (line-of-sight) velocity of clusters through the kinetic SZ effect (e.g., \citealp{Adam_2017, Sayers_2019}), complementing and establishing promising synergies in constraining the magnitude and distribution of turbulent flows in individual clusters.

In the meanwhile, numerical simulations of cosmological structure formation are amongst our best tools for exploring the physics of the ICM in order to interpret and lead future observations, as shown in the references above. A particularly relevant and fundamental issue is that of extracting a bulk and a turbulent part from an input velocity field (i.e., performing a Reynolds decomposition; \citealp{Adrian_2000}). This is especially challenging in the case of the ICM because of its strong multiscale character: while the outer accretion shocks which bound clusters can have curvature radii of a few Mpc, galactic processes, which have consequential effects for the whole cluster, occur on kpc scales or below. \cite{Vazza_2012} proposed an iterative algorithm for uniform grid data, further employed by \cite{Vazza_2017} and extended to SPH simulation outputs by \cite{Valdarnini_2019}, to constrain the outer scale of turbulence (roughly equivalent to the injection scale), which in turn allows to obtain a mean, bulk velocity field and a small-scale, turbulent velocity field. Recently, \citet{Shi_2018} presented a new approach, based on the wavelet decomposition of the velocity field, which allows not only to disentangle these two components, but also to construct a local spectrum of the velocity field.

Since the viscous scales are many orders of magnitude smaller than the resolution achievable by current, state-of-the-art simulations, small-scale fluctuations at the end of the energy cascade are dissipated by numerical viscosity. Several techniques have been applied to overcome this artifact and to extend the turbulent cascade, at least, to consistently model the relevant features of ICM turbulence: either constrained adaptive mesh refinement (AMR) simulations with an additional refinement scheme based on local vorticity ($\bm{\omega} = \nabla \times \vb{v}$) or velocity jumps \citep{IapichinoNiemeyer_2008, Vazza_2009, Vazza_2011}, or explicit subgrid models for the unresolved part of the turbulent cascade (e.g., \citealp{Schmidt_2016}; see also \citealp{Schmidt_2015_rev} for a review). Other authors have used uniform grids (or static refinement techniques which, ultimately, lead to a virial volume resolved within a fine, fixed grid; e.g., \citealp{Miniati_2014, Miniati_2015, Vazza_2017}). While these techniques have been successful in reproducing the expected properties of ICM turbulence, there are also evident downsides: for both, the refinement schemes based on vorticity or velocity jumps and subgrid models, one has to perform dedicated simulations, whose primary intent is to explore turbulent phenomena and which involve a much higher computational demand. As for static refinement simulations, these techniques are typically applied to single objects, rather than to a whole, cosmological population of clusters.

Large samples of clusters need to be analysed in order to achieve statistically significant conclusions. For this aim, given the extremely high computational demand of performing high-resolution simulations including full physics (taking into account that phenomena associated with feedback processes, such as galaxy motions or AGN feedback, are important for correctly modelling ICM turbulence), it can be challenging to perform dedicated simulations of large cosmological volumes including subgrid models or ad hoc refinement criteria, or either performing resimulations of a large number of objects. 

In this paper, we perform an exploratory work over an AMR simulation without subgrid modelling nor ad hoc refinement strategies, but with a base grid resolution which allows to resolve the bulk of gas mass within the virial volume of clusters with resolutions of $\Delta x \sim 20 \, \mathrm{kpc}$ or better. For this aim, we have developed several algorithms that take full advantage of the multiresolution description of AMR simulations. Then, following our study of the accretion history of galaxy clusters in \cite{Valles_2020}, in this work we connect the evolution of several global and local indicators of turbulence to the assembly history of the objects, mainly focusing on solenoidal turbulence, and we look into the mechanisms for its generation and dissipation.

The rest of the manuscript is organized as follows. In Sec. \ref{s:methods}, we describe our main numerical tools for analysing the velocity field in high-resolution simulation outputs. We detail the main features of our simulation, and the objects within it that will constitute the primary focus of this work, in Sec. \ref{s:simulation}. Our main results are then divided into Sec. \ref{s:global} and Sec. \ref{s:local}, where we present, respectively, a global and a local description of turbulence in the ICM. Last, we further discuss several aspects of our results in Sec. \ref{s:discussion} and present our conclusions in Sec. \ref{s:conclusions}. Appendix \ref{s:appendix.resolutionconvergence} discusses the convergence of volume-averaged quantities in pseudo-Lagrangian AMR sampled outputs.

\section{Methods}
\label{s:methods}
In order to provide a multiscale characterization of the properties of the three-dimensional velocity field, we have designed and implemented several algorithms that take full advantage of the AMR description. In Sec. \ref{s:methods.HHD}, we briefly describe our algorithm for splitting the velocity field in its rotational and compressive components. In Sec. \ref{s:methods.filter}, we present our filtering strategy to disentangle the bulk motions from the purely turbulent velocity field.

\subsection{Helmholtz-Hodge decomposition in an AMR multigrid}
\label{s:methods.HHD}
In order to split the velocity field in its compressive and solenoidal components, we perform a Helmholtz-Hodge decomposition (HHD). Our implementation, presented in \cite{Valles_2021_cpc}, does so by solving an elliptic equation for the scalar potential, $\phi$, which generates the compressive velocity, $\vb{v}_\mathrm{comp} = - \nabla \phi$; and one elliptic equation for each cartesian component of the vector potential, $\vb{A}$, which is responsible for the solenoidal (or rotational) velocity component, $\vb{v}_\mathrm{rot} = \nabla \times \vb{A}$.

The elliptic equations are addressed using standard methods for solving Poisson's equation in AMR cosmological simulations: namely, fast Fourier transform (FFT; see, e.g., \citealp{Press_1992}) techniques for the base level\footnote{While we use FFT for the base level, taking advantage of the periodic boundary of cosmological simulations, this is not an imposition of the method. If non-periodic boundary conditions are given, the base level could also be addressed with iterative methods, as we do for the refinement levels.}, and successive overrelaxation (SOR; \citealp{Young_1954}) for the refinement patches. All the derivatives are computed using high-order stencils to damp their contamination due to high-frequency noise. A more complete description and a set of tests can be found in \cite{Valles_2021_cpc}.

\subsection{Multiscale filtering}
\label{s:methods.filter}
Splitting the velocity field into a bulk (or coherent) and an inherently turbulent (uncorrelated, or \textit{small-scale}) component, i.e. performing a Reynolds decomposition, can be fairly subtle due to the absence of a univoque definition of turbulence \citep{Adrian_2000}. An early approach for this aim, used in cosmological simulations (e.g., \citealp{Dolag_2005, Vazza_2009}), consisted of filtering out the turbulent motions by subtracting the local mean velocity, defined over a fixed spatial scale. Another widely used strategy (e.g., \citealp{Lau_2009}) uses the velocity dispersion in radial shells as a proxy to quantify the level of turbulence around a particular cluster. Although these algorithms are conceptually simple and allow to effectively split the velocity fields in a bulk and a turbulent component, their fixed lengths do not fully capture the essentially multiscale nature of turbulent phenomena. 

More recently, \cite{Vazza_2012, Vazza_2017} proposed an algorithm to iteratively constrain the outer scale\footnote{In turbulence theory, the outer scale, $L(\vb{x})$, is the spatial scale at which turbulent motions are injected, and it corresponds to the largest scale inside the inertial range. In this text, we shall refer to the \textit{outer scale} or \textit{injection scale} interchangeably. We refer the reader to \citet[\S 26-38]{Landau_1987} for an introductory text on hydrodynamic turbulence.} for turbulence, $L(\vb{x})$, which then allows to extract the purely turbulent velocity field, $\delta \vb{v} (\vb{x})$, without explicitly fixing a filter length. While in \cite{Vazza_2017} the authors use a fixed refinement technique, so that the virial volume of their cluster is simulated within a uniform grid of constant, $\Delta x \sim 20 \, \mathrm{kpc}$ resolution, here we aim to extend the algorithm to a purely AMR velocity field, which allows us to straightforwardly apply the filter to clusters extracted from full-cosmological simulations without the need of performing resimulations. While the `physical' steps are parallel to those of \cite{Vazza_2012}, the multigrid, multiresolution description of an AMR simulation prevents us from using several handy tools (e.g., boxcar averages, convolutions, etc.) in the original formulation and, consequently, also increases substantially the computational cost. The basic steps of our implementation are as follows:

\begin{itemize}
	\item For each volume element in the computational domain, at a given refinement level $\ell$, the outer scale length, $L(\vb{x})$ is initialized to $L_{n=0} = 3 \Delta x_\ell$, where $\Delta x_\ell$ is the cell size at level $\ell$ and $n$ represents the algorithm iteration.
	\item For each cell, at position $\vb{x}$, $L(\vb{x})$ grows until convergence is reached according to the following iterative scheme:
	\begin{itemize}
		\item The bulk velocity at $\vb{x}$, at the iteration $n$, is computed as
		\begin{equation}
			\left< \vb{v} \right>_n(\vb{x}) = \frac{\iiint_{V'_{n}} \rho(\vb{x'}) \vb{v} (\vb{x'}) \mathrm{d} V'}{\iiint_{V'_{n}} \rho(\vb{x'}) \mathrm{d} V'}=\frac{\sum_{V'_{n}} m_i \vb{v}_i}{\sum_{V'_{n}} m_i} 
		\end{equation}
		\noindent where $\rho(\vb{x'})$ is the gas density at $\vb{x'}$, $m_i$ is the mass of the volume element $i$ and $V'_{n}$ is the spherical volume consisting of all the cells at positions $\vb{x'}$ such that $|\vb{x'} - \vb{x}| \leq L_n$ (at the maximum available resolution at each position).
		\item The turbulent velocity field at $\vb{x}$, at the iteration $n$, is computed as $\delta \vb{v}_n(\vb{x}) = \vb{v}(\vb{x}) - \left< \vb{v} \right>_n(\vb{x})$.
		\item Unless a stopping condition is triggered (see below), $L(\vb{x})$ is increased according to
		\begin{equation}
			L_{n+1} = \max\left(L_n + \Delta x_\ell, \, \left[1+\chi\right] L_n\right),
		\end{equation}
		\noindent where we fix $\chi = 0.05$. This condition, which has been tested experimentally, prevents a slow convergence when a high-resolution region has $L(\vb{x}) \gg \Delta x_\ell$.
	\end{itemize}
	\item For each cell, the iterative scheme is stopped whenever at least one of the following conditions is met:
	\begin{itemize}
		\item The fractional change in $\delta \vb{v}$ between two consecutive iterations, which we define as 
		\begin{equation}
			\Delta = \max_{i=1,2,3} \left|\frac{\delta v_n^i}{\delta v_{n-1}^i} - 1 \right|,
		\end{equation}
		falls beyond a fixed tolerance parameter, $\Delta \leq \Delta_\mathrm{tol}$. When this condition is achieved, $L_n$ represents the maximum correlation scale of the velocity field at $\vb{x}$. We have set $\Delta_\mathrm{tol} = 0.05$, although in our tests the turbulent velocity field does not depend strongly on this threshold, as it was also reported by \cite{Vazza_2012} and \cite{Valdarnini_2019}.
		\item A shocked cell, which we define as a cell with Mach number $\mathcal{M} \geq 1.3$ (e.g. \citealp{Quilis_1998, Ryu_2003, Vazza_2009_shocks, Planelles_2013, Martin-Alvarez_2017}), enters the volume $V'_{n}$. We detect shocks with the shock finder presented in \cite{Planelles_2013}, which uses the one-dimensional discontinuities in temperature to solve for $\mathcal{M}$. The shock exclusion is well motivated by the fact that shocks introduce velocity discontinuities, which importantly bias the bulk velocity determination.
		\item The volume $V'_{n}$ intersects the computational domain's boundary. Even though we could explicitly enforce the periodic boundary conditions to get rid of this limitation, this condition only biases the results in a negligible amount of volume close to the boundary, where we do not expect to find objects of interest.
	\end{itemize}
\end{itemize}

Once this procedure is repeated for all the volume elements, we have a complete, multiscale description of the coherence scale, $L(\vb{x})$, and the bulk and turbulent\footnote{Through this manuscript, we shall also refer to $\delta \vb{v} (\vb{x})$ as the \textit{filtered} or the \textit{small-scale} velocity field.} velocity fields. Naturally, the turbulent velocity field can also be decomposed into its rotational and compressive components using the algorithm described in Sec. \ref{s:methods.HHD}, yielding the solenoidal and compressive turbulent velocity fields, respectively. The routines for applying this filtering scheme have been included in the publicly available code \texttt{vortex}\footnote{\url{https://github.com/dvallesp/vortex}} \citep{Valles_2021_cpc}.

\section{The simulation}
\label{s:simulation}

The remaining pages of this manuscript are based on the results of the codes described in Sec. \ref{s:methods} applied to a high-resolution cosmological simulation, which is described below. In Sec. \ref{s:simulation.structurefinding}, we outline the structures which constitute the primary focus of this work.

The simulation analysed in this paper has been carried out with \textsc{MASCLET}, an Eulerian AMR, \textit{high-resolution shock-capturing} hydrodynamics, coupled to particle-mesh $N$-body, cosmological code \citep{Quilis_2004}. This same simulation has been employed in several previous works \citep{Quilis_2017, Planelles_2018, Valles_2020}. Here, we describe the main features of the simulation. For more information about some details not directly connected to our analyses, we refer the reader to the aforementioned references.

The simulation corresponds to a cubic, periodic domain of comoving side length $40 \, \mathrm{Mpc}$, with a flat $\Lambda$CDM cosmology set up by a Hubble parameter $h \equiv H_0 /( 100 \, \mathrm{km\, s^{-1}\,Mpc^{-1}}) = 0.678$, composition given by the matter, baryon and dark energy parameters $\Omega_m = 0.31$, $\Omega_b = 0.048$, $\Omega_\Lambda = 0.69$, and primordial fluctuation spectrum set according to a spectral index $n_s = 0.96$ and amplitude $\sigma_8 = 0.82$. Thus, the cosmology is consistent with the latest values reported by the \cite{Planck_2018_VI}. The domain has been discretized in a base grid of $128^3$ cells, which yields a harsh resolution of $\sim 310\, \mathrm{kpc}$. Regions with large gaseous and/or DM densities get recursively refined following a pseudo-Lagrangian AMR scheme with up to $n_\ell = 9$ refinement levels, providing a peak resolution of nearly $\sim 610\, \mathrm{pc}$. With four species of DM particles, the best DM mass resolution is $\sim 2 \times 10^6 M_\odot$, equivalent to filling the domain with $1024^3$ of such particles.

The initial conditions were set up by a CDM transfer function \citep{Eisenstein_1998} at redshift $z=100$. A constrained realization of the gaussian random field, according to the procedure of \cite{Hoffman_1991}, was used to produce a massive cluster in the centre of the box. Besides gravity, the simulation accounts for several cooling mechanisms (free-free, inverse Compton and atomic and molecular cooling for a primordial gas), as well as heating by a UV background of radiation \citep{Haardt_1996}. Star formation and type-II supernova feedback are parametrized according to \cite{Yepes_1997} and \cite{Springel_2003}. Even though this run does not include AGN feedback, which could constitute an important source of turbulent motions in the innermost regions of galaxy clusters (see, e.g.,\citealp{Vazza_2012}), this drawback is not especially relevant for this work, where our primary intent is performing an exploratory analysis showing the capabilities of our algorithms. We may postpone further analyses, including AGN feedback and magnetic fields, to a future work.

\subsection{Structure finding}
\label{s:simulation.structurefinding}
We have identified the structures in this computational domain by means of the spherical overdensity DM halo finder \textsc{ASOHF} \citep{Planelles_2010, Knebe_2011}. At $z\simeq 0$, there are two massive galaxy clusters (with virial masses $M_\mathrm{vir,DM} > 10^{14} M_\odot$), which we shall hereon refer to as CL01 and CL02.

\begin{table}
	\centering
	\small
	\caption{Main properties of clusters CL01 and CL02 at $z=0$. Virial radii, $R_\mathrm{vir}$, are defined with respect to the DM distribution, according to the standard spherical overdensity definition \citep{Lacey_1994} with the virial overdensity given by \citealp{Bryan_1998}. $M_\mathrm{DM}$ and $M_\mathrm{gas}$ are measured inside $R_\mathrm{vir}$. Temperatures and entropies are computed inside $R_\mathrm{vir}$, assuming hydrostatic equilibrium \citep[equations 59 and 64]{Voit_2005} and with a mean molecular weight $\mu = 0.6$.}
	\begin{tabular}{c|ccccc}
		\hline
		cluster & $R_\mathrm{vir}$ & $M_\mathrm{DM}$ & $M_\mathrm{gas}$ & $k_B T_\mathrm{vir}$ & $K_{e,\mathrm{vir}}$ \\
		& (Mpc) & ($10^{13} M_\odot$) & ($10^{13} M_\odot$) & (keV) & ($\mathrm{keV \, cm^2}$) \\
		\hline
		CL01 & 1.99 & 42.9 & 4.56 & 3.27 & 1230 \\
		CL02 & 1.26 & 10.9 & 1.33 & 1.34 & 520 \\
		\hline
	\end{tabular}
	\label{tab:cluster_properties}
\end{table}

In \cite{Valles_2020}, we have analysed the accretion histories of these two objects. CL01 is a massive cluster which suffers several major and minor merging periods through its recent history, the most recent major merger having occurred at around $z\sim 0.9$ (see figure 2 in \citealp{Valles_2020}), and only experiencing quiescent accretion therein, with low accretion rates from $z \sim 0.4$ on. Likewise, CL02 is a smaller, $\sim 10^{14} M_\odot$ cluster which has experienced a major merger at $z \sim 1.4$ and a minor merger at around $z \sim 0.2$, the latter not having any significant impact on the cluster's structure and, thus, probably having had reduced dynamical relevance, as far as turbulence is concerned. A summary of the properties of these two objects, at $z \simeq 0$, is shown in Table \ref{tab:cluster_properties}.

\begin{figure}
\centering
\includegraphics[width=\linewidth]{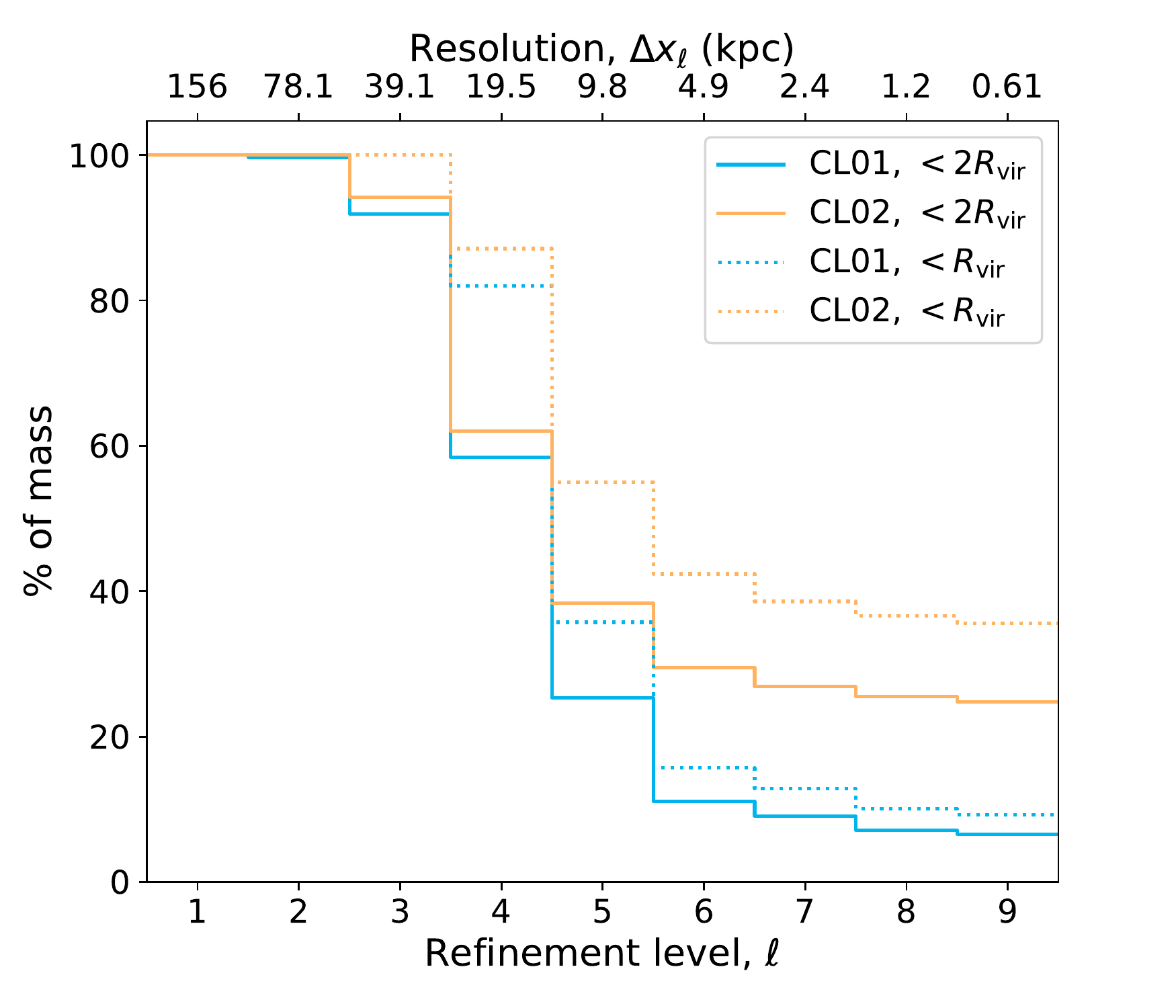}
\caption{Description of the resolution coverage of each cluster (within $2 R_\mathrm{vir}$, solid lines; and within $R_\mathrm{vir}$, dotted lines) at $z \simeq 0$. For each refinement level, $\ell$, or resolution, $\Delta x_\ell$, the vertical axis indicates the fraction of mass inside the considered volume in cells with resolution $\Delta x_\ell$ or better.}
\label{fig:resolution_CL01_CL02}
\end{figure}

As the description of the turbulent flows is strongly dependent on the resolution of the numerical grid, we present the (cumulative) fraction of gas mass (within $2R_\mathrm{vir}$, solid lines; and $R_\mathrm{vir}$, dashed lines) resolved, by $z \sim 0$, at each refinement level in Figure \ref{fig:resolution_CL01_CL02}. For both clusters, $\gtrsim 60\%$ of the mass inside $2 R_\mathrm{vir}$ is resolved in patches with resolution at least $\Delta x_4 \sim 20 \, \mathrm{kpc}$. This resolution can be directly compared to previous works which employ ad hoc resimulations and fixed refinement techniques (e.g., \citealp{Vazza_2017}, who resolve the virial volume of their clusters within a uniform grid with an equivalent resolution; see also \citealp{Miniati_2014, Miniati_2015}, who achieve a $\sim 10 \, \mathrm{kpc}$ resolution throughout the virial volume of their cluster).

If we restrict to the virial volume, virtually all the mass is resolved at resolutions equal or better than $\sim 40 \, \mathrm{kpc}$, and $\gtrsim 80\%$ at resolutions at least $\sim 20 \, \mathrm{kpc}$. Additionally, almost $10\%$ (CL01) and $40\%$ (CL02) of the virial mass is resolved with resolutions below the $\mathrm{kpc}$.

\section{Global statistics of turbulence}
\label{s:global}
In this section, we analyse several global indicators of turbulence. In particular, we explore the second-order velocity structure functions (Sec. \ref{s:global.strfunc}), their evolution (at fixed scales) in relation to accretion rates and merger events (Sec. \ref{s:global.evolution}) and the scaling relations and mechanisms for the generation of solenoidal turbulence (Sec. \ref{s:global.vorticity_helicity}).

\subsection{Turbulence spectra: structure functions}
\label{s:global.strfunc}
Structure functions quantify the magnitude of the velocity fluctuations on different scales over the cluster volume (or mass), and therefore offer a straightforward way to define a global statistic of turbulence for a given object. The most direct definition of the structure function of order $p$, $S_p(L)$, is given by the expression

\begin{equation}
	S_p (L) = \left< |\vb{v}(\vb{x} + L \vb{\hat n}) - \vb{v}(\vb{x})|^p \right>_{\vb{x},\vb{\hat{n}}},
	\label{eq:def_structure_function}
\end{equation}

\noindent where $\vb{v}$ will hereon denote the (\textit{unfiltered}, i.e. total) peculiar velocity field, $\vb{\hat n}$ is a unit vector and the average is carried both in positions, $\vb{x}$, inside the volume of interest and in directions, $\vb{\hat n}$. Using the HHD algorithm presented in \cite{Valles_2021_cpc} (see Sec. \ref{s:methods.HHD}), we will also compute these structure functions separately for the compressive and solenoidal velocity components. While some authors (e.g., \citealp{Valdarnini_2011, Miniati_2014}) further decompose the structure functions in a longitudinal and a transverse component, we shall not pursue such decomposition here, as we are only interested in the magnitudes of the velocity fluctuations at each scale.

In order to correctly ponder the dynamically relevant regions, we adopt a mass weight when performing the average over $\vb{x}$. In particular, our procedure for computing $S_p(L)$ can be summarised in the following steps:

\begin{enumerate}
	\item We choose $N_s$ mass-weighted random points (hereon, the \textit{sampling points}) inside the considered volume (in our case, inside a sphere twice the virial radius around each cluster). The weighted sampling is performed by applying Smirnov's inverse transformation method (see, e.g., \citealp{Devr86}), i.e, we compute the cell-wise cumulative normalized mass distribution inside $2 R_\mathrm{vir}$, draw uniformly sampled random numbers, $s \in [0,1]$, and apply the inverse cumulative function to these numbers. 
	
It is worth noting that gas clumps and substructures can importantly bias the random selection of points, especially in this simulation as the lack of AGN feedback could produce overcooling (see, e.g., \citealp{Kravtsov_2005, Eckert_2012, Zhuravleva_2013, Planelles_2014}). In order to avoid our sampling points to concentrate in highly overdense gas clumps, whose dynamics may differ from those of the bulk ICM, we first apply the inhomogeneity identification technique described by \cite{Zhuravleva_2013}. In particular, we radially split the cluster in $N_\mathrm{bins} = 20$ logarithmic bins\footnote{We note that the substructure-excised density profiles are not sensitive to the particular choice of $N_\mathrm{bins}$ in the range $N_\mathrm{bins} \in [15,100]$.}, and tag as clumps all the cells whose logarithmic density is further than $f_\mathrm{cut} = 3.5$ standard deviations from the volume-weighted median log-density. 
	\item Around each of the sampling points, we pick $N_f$ volume-weighted random points (hereon, the \textit{field points}). In order to effectively sample all the scales, we extract the field points around each sampling point, $\vb{x}_s$, using the same method as above but weighting the probability of each cell by $V_\mathrm{cell} / |\vb{x}_s - \vb{x}_\mathrm{cell}|^3$. It is easy to show that this weight ensures equal number of points per logarithmic radial bin.
	\item Finally, we bin logarithmically the distances among points, $L$, and average over all the possible pairs at each bin according to expression (\ref{eq:def_structure_function}).
\end{enumerate}

In our tests, taking $N_s = 5000$ mass-weighted points is enough to adequately sample the mass distribution within a precision of $5\%$ or better. We take as well up to $N_f = 5000$ volume-weighted field points around each sampling point, yielding a maximum of $N_s \times N_f = 2.5 \times 10^7$ pairs of points to compute the structure function. We then bin $L$ in $50$ logarithmic intervals, so that the final value of $S_p(L)$ at each $L$ has been computed as the average over $\lesssim 5 \times 10^5$ pairs. In order to check whether these statistics are robust enough, we have performed 20 bootstrap iterations of these procedures, verifying a small variance in our results (typically in the order of $\sim 1\%$, and $\lesssim 10\%$ for any bin).

\begin{figure*}
\centering
{\includegraphics[width=0.5\linewidth]{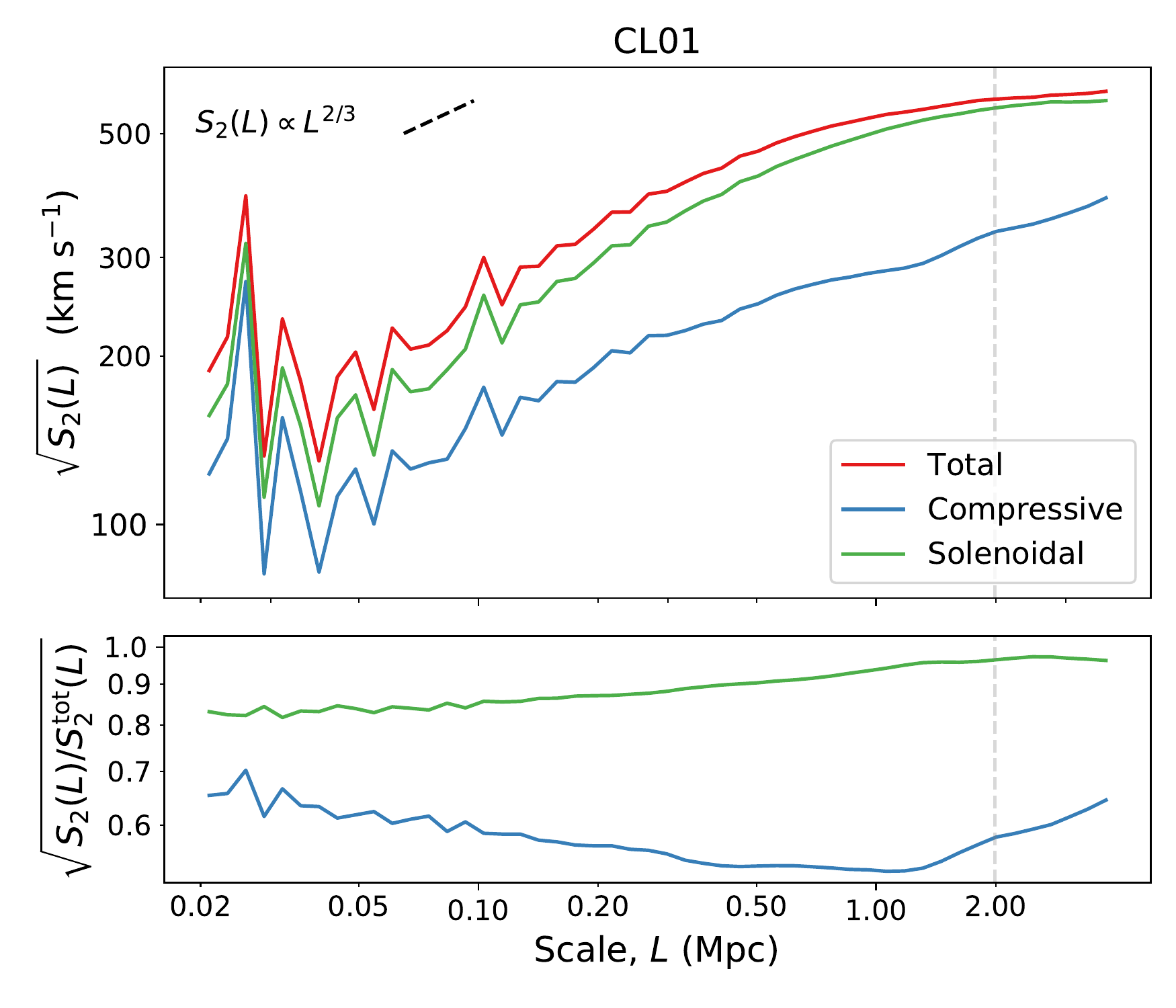}}~
{\includegraphics[width=0.5\linewidth]{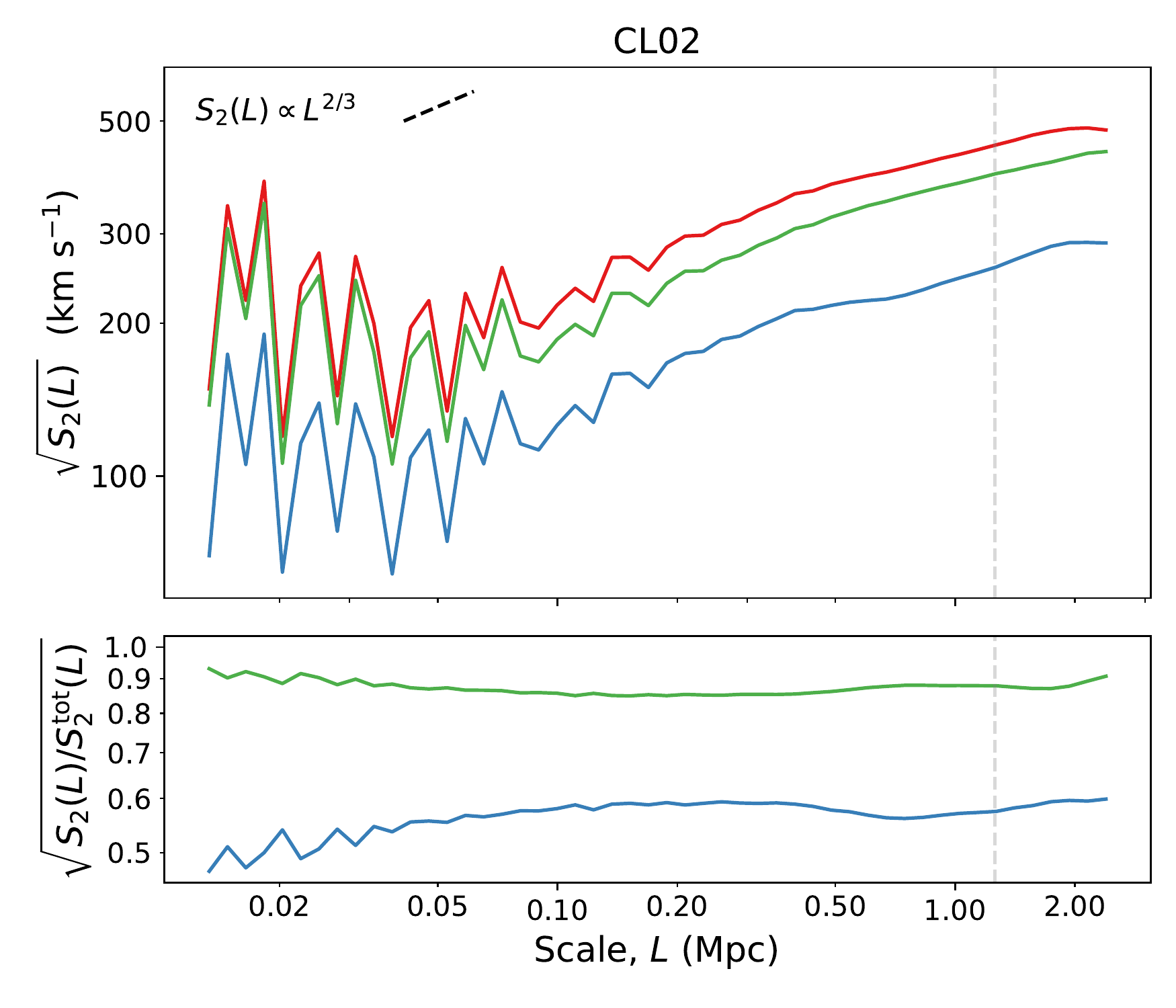}}
\caption{Second-order structure functions for clusters CL01 (left) and CL02 (right) at $z \simeq 0$. The red, green and blue lines correspond to the total, solenoidal and compressive velocity fields, respectively. The gray, dashed, vertical lines mark the scale of the virial radii. The black, dashed lines indicate the slope expected by the Kolmogorov scaling \citep{Kolmogorov_1941}. The square root of the second-order structure functions is represented to show the results in units of velocity.}
\label{fig:structure_functions}
\end{figure*}

We present in Figure \ref{fig:structure_functions} the second-order ($p=2$) structure functions for the total velocity, as well as for the compressive and the rotational velocity components, for clusters CL01 (left-hand panel) and CL02 (right-hand panel) at $z \simeq 0$. For comparison, in both panels we show the slope predicted by the \cite{Kolmogorov_1941} model for fully developed isotropic, incompressible turbulence, $S_2(L) \propto L^{2/3}$. Both components (as well as the total velocity) present a well-defined inertial range, whose slope roughly agrees with the Kolmogorov prediction. The structure functions for the solenoidal component are consistently higher than their compressive counterparts, mainly as a consequence that solenoidal flows are much more ubiquitous and volume filling. However, at variance with previous results with static refinements, we do not find a clear difference in the slopes of these components. \cite{Miniati_2014, Miniati_2015} finds the compressive component to have a steeper spectrum, the opposite trend being reported by \cite{Vazza_2017}. The lower panels in Fig. \ref{fig:structure_functions} show the quotient of the solenoidal and compressive structure functions to the total one. While CL01 presents a higher slope in the solenoidal component, the opposite happens for CL02. This highlights that these slopes could be particularly sensitive to the dynamical state of the cluster, as well as to the numerical scheme and the sampling of $S_p(L)$.

The structure functions flatten at $L \gtrsim 1 \, \mathrm{Mpc}$, placing the characteristic scale for turbulence injection at around $(0.5-1)R_\mathrm{vir}$, in consistence with \cite{Miniati_2014} and \cite{Vazza_2017}\footnote{Other works (e.g., \citealp{Vazza_2012}) report slightly smaller injection scales, in the range $(0.1-0.3)R_\mathrm{vir}$. In any case, turbulence is not expected to be injected by bulk motions from any specific length, but from a wide range of scales instead.}. The energy that bulk motions inject into the turbulent velocity field then cascades down to smaller scales through different instabilities (e.g., Kelvin-Helmholtz; see for example \citealp{ZuHone_2011}), until they get (numerically) dissipated when the eddy sizes are in the order of several times the cell side length. The pseudo-Lagrangian refinement approach of our AMR implementation implies that different mass elements in the cluster get different dissipation scales, originating oscillations in the small-scale end of the structure functions. These oscillations are also present in the `Lagrangian AMR' runs of \cite{Miniati_2014}, and limit the extent to which structure functions are a useful statistic to study turbulence in full-cosmological simulations using pseudo-Lagrangian approaches.

\subsubsection{Characteristic time-scales of turbulent eddies}
\label{s:global.strfunc.timescales}
The time-scale in which eddies of size $L$ cascade down to the dissipation scale can be estimated by the so-called \textit{eddy turn-over time}, $\tau_\mathrm{eddy} \sim L / \mathcal{V}$, where $\mathcal{V}$ is a characteristic velocity at the scale $L$. In the framework of the \cite{Kolmogorov_1941} model, this quantity is expected to scale as $\tau_\mathrm{eddy} \propto L^{2/3}$. It is customary through the bibliography (e.g., \citealp{Miniati_2014}) to just choose a length ($L \sim R_\mathrm{vir}$) and a characteristic velocity representing the overall cluster dynamics ($v_\mathrm{vir} \equiv \sqrt{G M_\mathrm{vir}/R_\mathrm{vir}}$) to have a time-scale estimate. Even though these can only be taken as order-of-magnitude estimates, we argue that a more self-consistent estimation can be got by using the $S_2(L)$ spectrum, as $\tau_\mathrm{eddy} (L) \sim L / \sqrt{S_2(L)}$, since it is a property intrinsic to the gas velocity field (and not of the gravitational potential well, as the one derived from the circular velocity inside a clustercentric radius $L$, $v_\mathrm{circ}(L)$).

\begin{figure}
\centering
{\includegraphics[width=\linewidth]{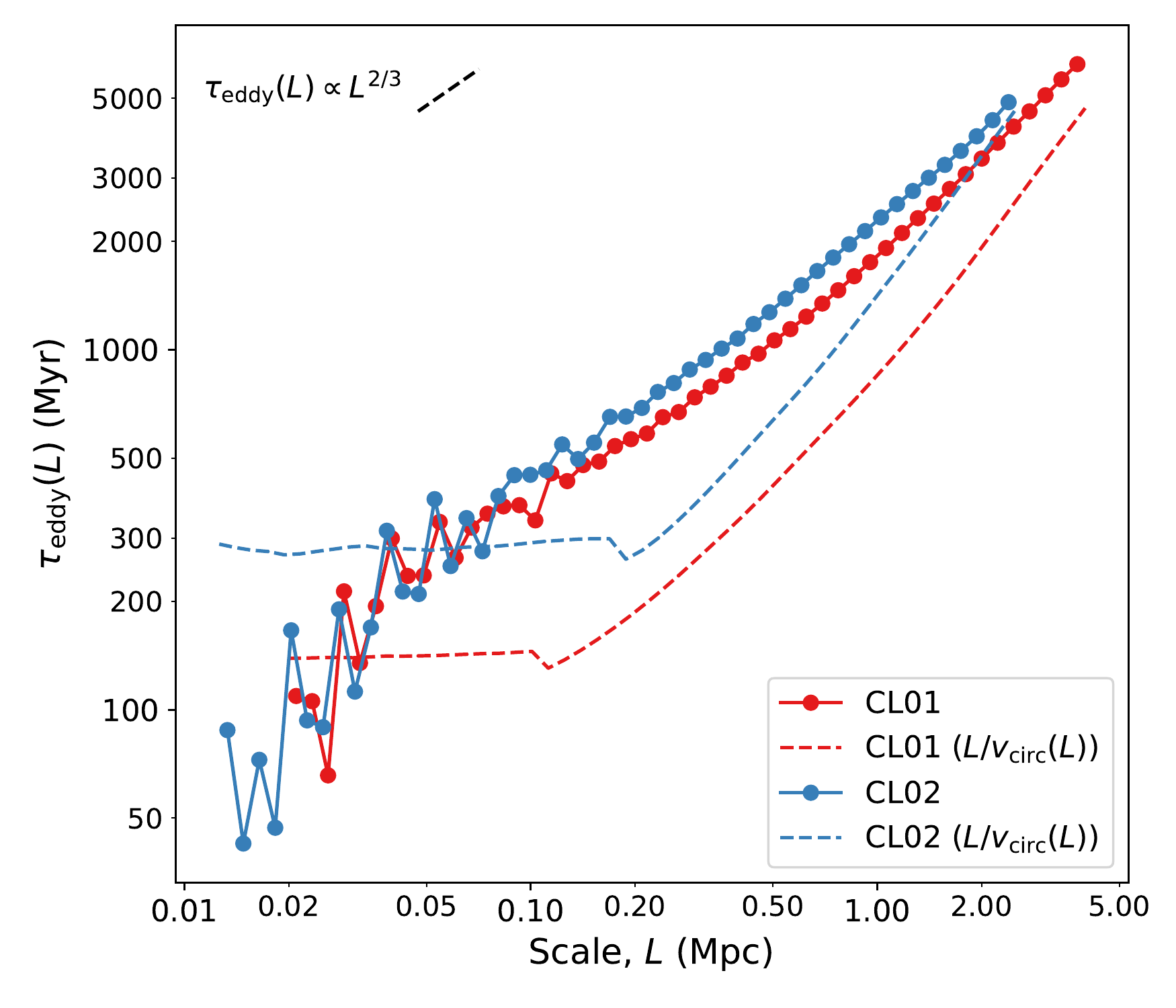}}
\caption{The dots joined by solid lines show the eddy turn-over times for different scales, $\tau_\mathrm{eddy} (L)$, estimated using the structure functions shown in Fig. \ref{fig:structure_functions} for cluster CL01 (red lines) and CL02 (blue lines). For comparison, the dashed lines show the time-scales estimated from the circular velocities inside spheres $r<L$ around the cluster centre.}
\label{fig:timescales}
\end{figure}

Figure \ref{fig:timescales} presents the eddy turn-over time-scales for clusters CL01 and CL02 (red and green dots, respectively) at redshift $z \simeq 0$. The time-scales for both clusters are remarkably similar through all the scale range, despite the difference in masses. Across the inertial range, $\tau_\mathrm{eddy}(L)$ scales in good agreement with the Kolmogorov prediction. For comparison, we also show, in dashed lines, the times-cale derived from the circular velocity inside a radius $L$. Assuming this magnitude to be dominated by the DM mass distribution, and taking an NFW \citep{NFW_97} profile for DM density, it is easy to show that the time-scale computed that way should scale as $L^{3/2} / \log (L) \sim L^{3/2}$ for sufficiently large $L$ ($L \gg r_s$, being $r_s$ the scale radius of the NFW profile). Indeed, this is the behaviour observed in our clusters. Interestingly, these two time-scales seem to approach for $L \gtrsim R_\mathrm{vir}$, making $R_\mathrm{vir} / v_\mathrm{vir}$ an incidentally good guess for the characteristic time-scale of the turbulent energy cascade for injection scales comparable to the virial radius. However, we emphasize that the time-scale $\tau_\mathrm{eddy} = L / \sqrt{S_2(L)}$ better captures the multiscale nature of turbulent motions.

By fitting the inertial ranges of the timescale curves to a power law, for each of the clusters, we get the following fitting formula for the cascade timescales of eddies of size $L$:

\begin{equation}
	\tau_\mathrm{eddy}(L) \approxeq (1755\pm78) \, \mathrm{Myr} \left(\frac{L}{1\, \mathrm{Mpc}}\right)^{0.647\pm0.026} 
	\label{eq:fit_timescales}
\end{equation}

Naturally, these fitting formulae have to be taken with caution, since a proper statistical analysis, which we may defer for a future work, must be performed in order to evaluate its universality. Nevertheless, within the context of this work and focusing at recent redshifts, we shall use this expression to interpret the results in the next sections.

\subsubsection{Behaviour of the different phases}
\label{s:global.strfunc.phases}

\begin{figure*}
\centering
{\includegraphics[width=0.5\linewidth]{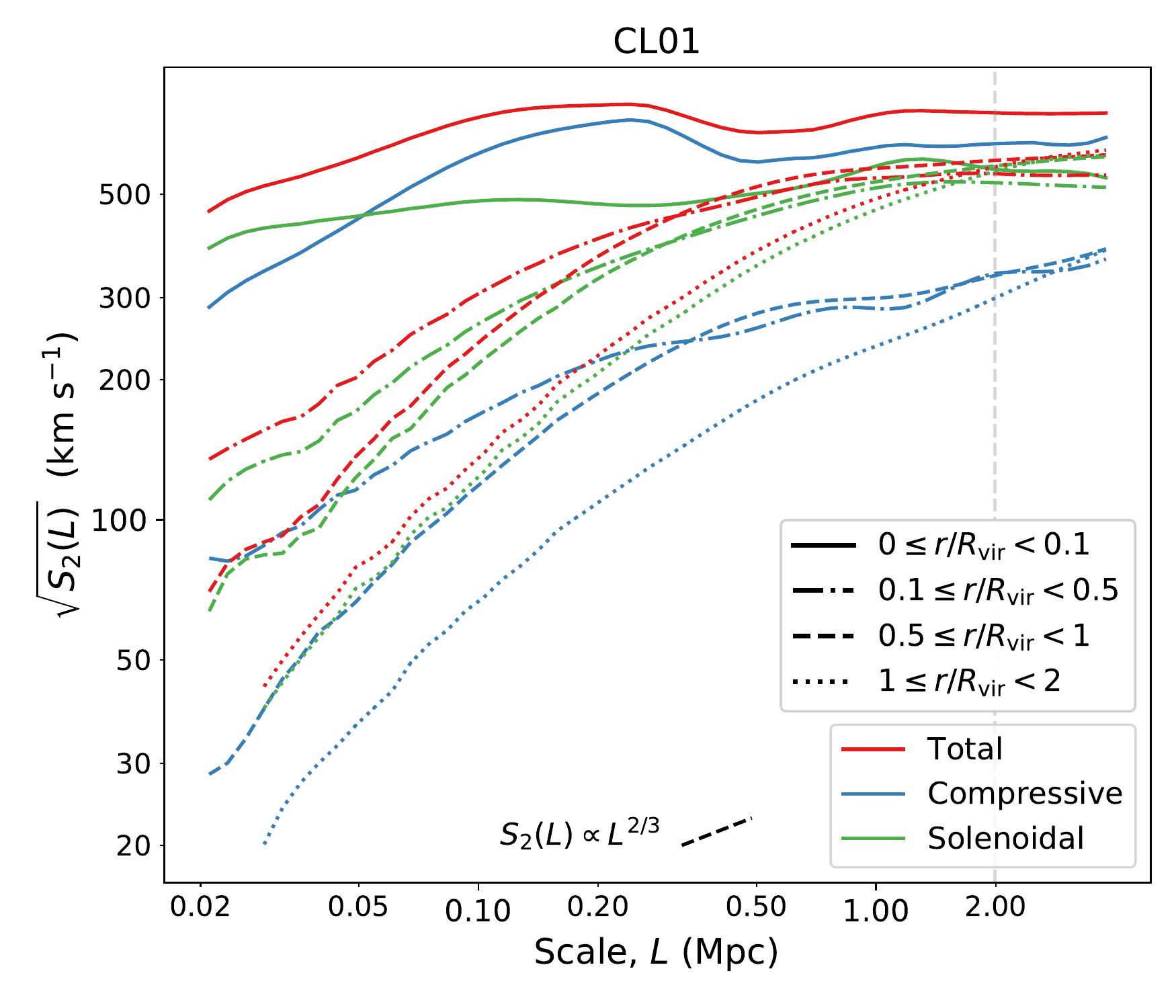}}~
{\includegraphics[width=0.5\linewidth]{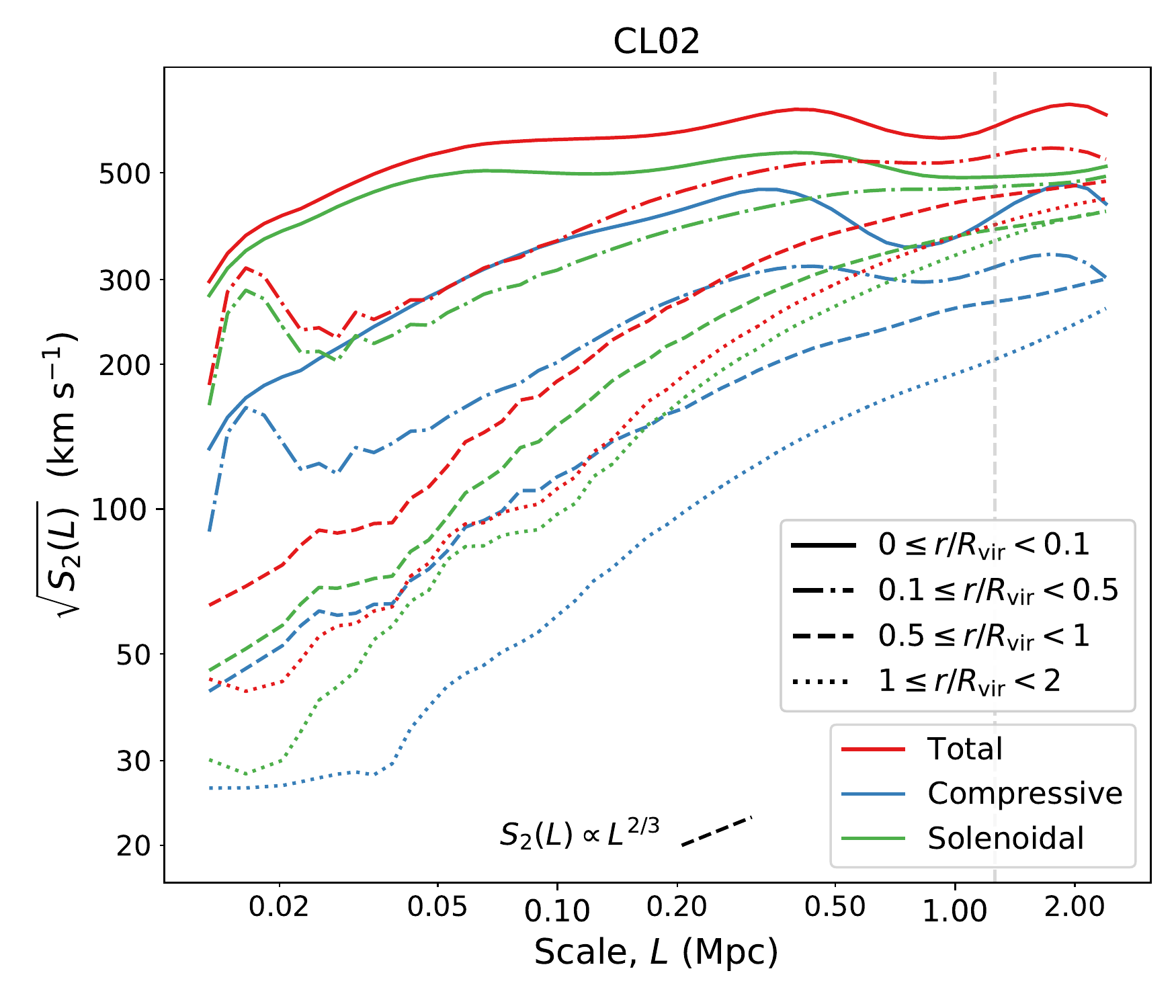}}
\caption{Second-order structure functions for different regions of clusters CL01 (left) and CL02 (right) at $z \simeq 0$. Solid, dot-dashed, dashed and dotted lines correspond to the core, off-core, virial and outskirts regions, as defined in Sec. \ref{s:global.strfunc.phases}. The rest of notations are the same as in Fig. \ref{fig:structure_functions}.}
\label{fig:structure_functions_phases}
\end{figure*}

Although the average statistics of the ICM velocity field show general properties which roughly agree with the Kolmogorov scaling, the behaviour can vary significantly for different phases. While some authors perform a phase-space splitting and study the properties of the velocity fields of the different phases (e.g., \citealp{Schmidt_2016}), here we find more clear and reproducible to take a geometric criterion to study the dependence of the velocity structure functions. Thus, we consider the following regions: \textit{core} ($0 < r/R_\mathrm{vir} < 0.1$), \textit{off-core} ($0.1 < r / R_\mathrm{vir} < 0.5$), \textit{virial} ($0.5 < r / R_\mathrm{vir} < 1$) and \textit{outskirts} ($1 < r / R_\mathrm{vir} < 2$). Similar choices are made in previous works (e.g., \citealp{Miniati_2014}). For each phase, we only count the sampling points inside the considered region, while, naturally, field points can be in any position within the $2R_\mathrm{vir}$ volume. The results for CL01 and CL02 are presented in Figure \ref{fig:structure_functions_phases}. In this case, we have smoothed the functions using a \cite{SavitzkyGolay_1964} filter to alleviate the oscillations, whose origin has already been commented, and focus on the general scaling.

For both clusters, at large clustercentric distances, the small-scale velocity fluctuations are strongly suppressed, as these regions are typically resolved within refinement levels $\ell = 2$ or $\ell = 3$ (thus, with an effective resolution of $\sim 40 \, \mathrm{kpc}$ or worse, and placing the numerical dissipation scale at several hundreds of $\mathrm{kpc}$), and, consequently, the structure functions are much steeper than Kolmogorov's. Although these trends do vary significantly from cluster to cluster (and with cosmic time), this lack of small-scale power, especially for the solenoidal component, is also reported by \cite{Miniati_2014} in their `Lagrangian AMR' run. 

On the other hand, cluster cores are extremely sensitive to the feedback mechanisms which are accounted for in the simulation (see, for example, \citealp{Rasia_2015} and \citealp{Planelles_2017}). In this case, the lack of AGN feedback prevents the injection of random motions in the innermost regions of clusters. Despite this effect, fluctuations on these inner regions are considerably higher than in the rest of regions, both because of numerical (these regions are likely to be mostly resolved with cell sizes of $\lesssim \mathrm{kpc}$, placing the dissipation scale below $\sim 10 \mathrm{kpc}$) and physical (SNe feedback and inner, merger shocks can be important sources of turbulent motions) reasons. \cite{Valdarnini_2011, Valdarnini_2019} suggests a scenario where small-scale turbulence is generated in the cluster central regions due to the interaction of a dense, compact core with the surrounding ICM in runs with cooling but not central sources of energy. This could, indeed, be the case of our clusters. Both CL01 and CL02 have flat density profiles in the central $\sim 100 \, \mathrm{kpc}$ and lower temperatures in their cores than the surrounding ICM, thus corresponding to cool, compact cores \citep{Burns_2008}. Therefore, this scenario is capable of providing a plausible explanation for the flattening of the structure functions in inner radii. 

\subsection{Evolution of the velocity fluctuations on different scales}
\label{s:global.evolution}
As seen through Sec. \ref{s:global.strfunc}, the global statistics of the velocity field in pseudo-Lagrangian AMR simulated galaxy clusters present systematic deviations from those obtained in other works using fixed grids. However, we shall argue that these differences emerge noticeably wherever the resolution is low (i.e., in low-density regions such as cluster outskirts). Even though our clusters simulated in a full-cosmological environment can lack small-scale velocity power in a fraction of their volume, the \textit{ansatz} behind the AMR strategy based on local density implies that this has indeed a small contribution to the total energetics and dynamical evolution of the cluster, since the mass fraction corresponding to this large volume is generally reduced.

As mentioned in Sec. \ref{s:global.strfunc.phases}, the actual behaviour of the structure functions is highly dependent on the dynamical state, and thus it is interesting to investigate its time evolution. In order to avoid the systematic effects seen in the velocity structure functions due to the AMR grid structure, we will focus on the values of the structure function for certain pre-defined scales, large enough to neglect the effect of the aforementioned oscillations. Hereon we will refer to these quantities as the \textit{velocity fluctuations at the scale $L$}. Thus, even though the slope of our structure functions could be biased with respect to uniform grid runs, a large part of these systematics is cancelled out by focusing on constant scales, and the evolution of the velocity fluctuations can be properly assessed. 

In particular, we compute $S_2(L)$ through each of the 41 code outputs from $z \simeq 1.5$ to $z \simeq 0$ at the scales $L/R_\mathrm{vir} = 0.1,\, 0.5,\, 1,\, \text{and} \, 1.5$. The procedure is identical to the one described in Sec. \ref{s:global.strfunc}, but in this case we only generate uniformly distributed field points at radial distances corresponding to the four considered scales around each sampling point (with a given tolerance, which we have fixed to $10\%$).

\subsubsection{Relation to accretion and mergers}
\label{s:global.evolution.accretion}
Accretion and mergers are important sources of energetic feedback to the ICM, and are directly connected to the dynamical state of the cluster (see, e.g., \citealp{Quilis_1998, Planelles_2009, Lau_2015, Chen_2019, Valles_2020}). In order to investigate their role in generating turbulence, we will correlate the velocity fluctuations on the scales defined above with the merger periods and the instantaneous accretion rates.

Following \cite{Diemer_2014}, we define the mass-accretion rate (MAR) as the logarithmic rate of change of the enclosed mass with respect to the scale factor. Operationally, and as opposed to other works which use an averaged MAR over a wide time interval, in \cite{Valles_2020} we defined an instantaneous MAR,

\begin{equation}
	\Gamma_\Delta(a) = \dv{\log M_\Delta}{\log a}
	\label{eq:mar_definition}
\end{equation} 

\noindent computed from the sparse snapshot sampling using \cite{SavitzkyGolay_1964} filters for performing the differentiation, in order to mitigate the contamination  of sampling noise on the numerical derivatives. For the analyses in this work, we will be interested in the baryonic (thus, $M_\Delta$ being the combined mass of gas and stars) and total (baryonic and dark matter) MARs.

We split the recent ($1.5 \gtrsim z \geq 0$) history of the clusters in three merging regimes, namely `ongoing major merger', `ongoing minor merger' and `smoothly accreting'. For classifying the mergers, we use the mass ratio of the two most massive progenitor clusters \citep{Planelles_2009}. Mergers above a 1:3 mass ratio are regarded as \textit{major mergers}. Those between 1:3 and 1:10 are tagged as \textit{minor mergers}. Mergers below the 1:10 threshold are not considered, and a cluster not experiencing any merger above this threshold at a given redshift is classified as smoothly accretting. 

\begin{figure*}
\centering
{\includegraphics[width=0.5\linewidth]{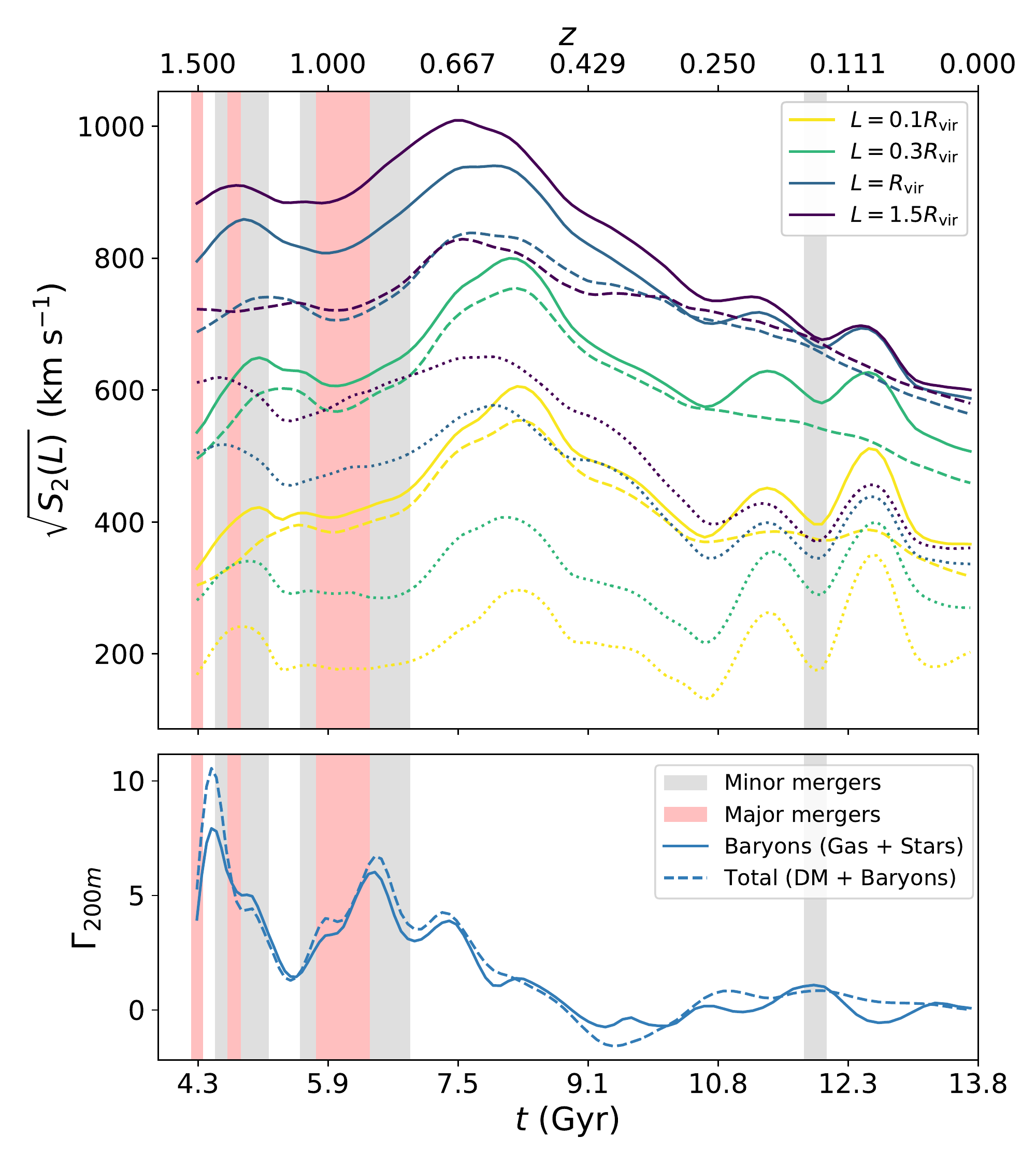}}~
{\includegraphics[width=0.5\linewidth]{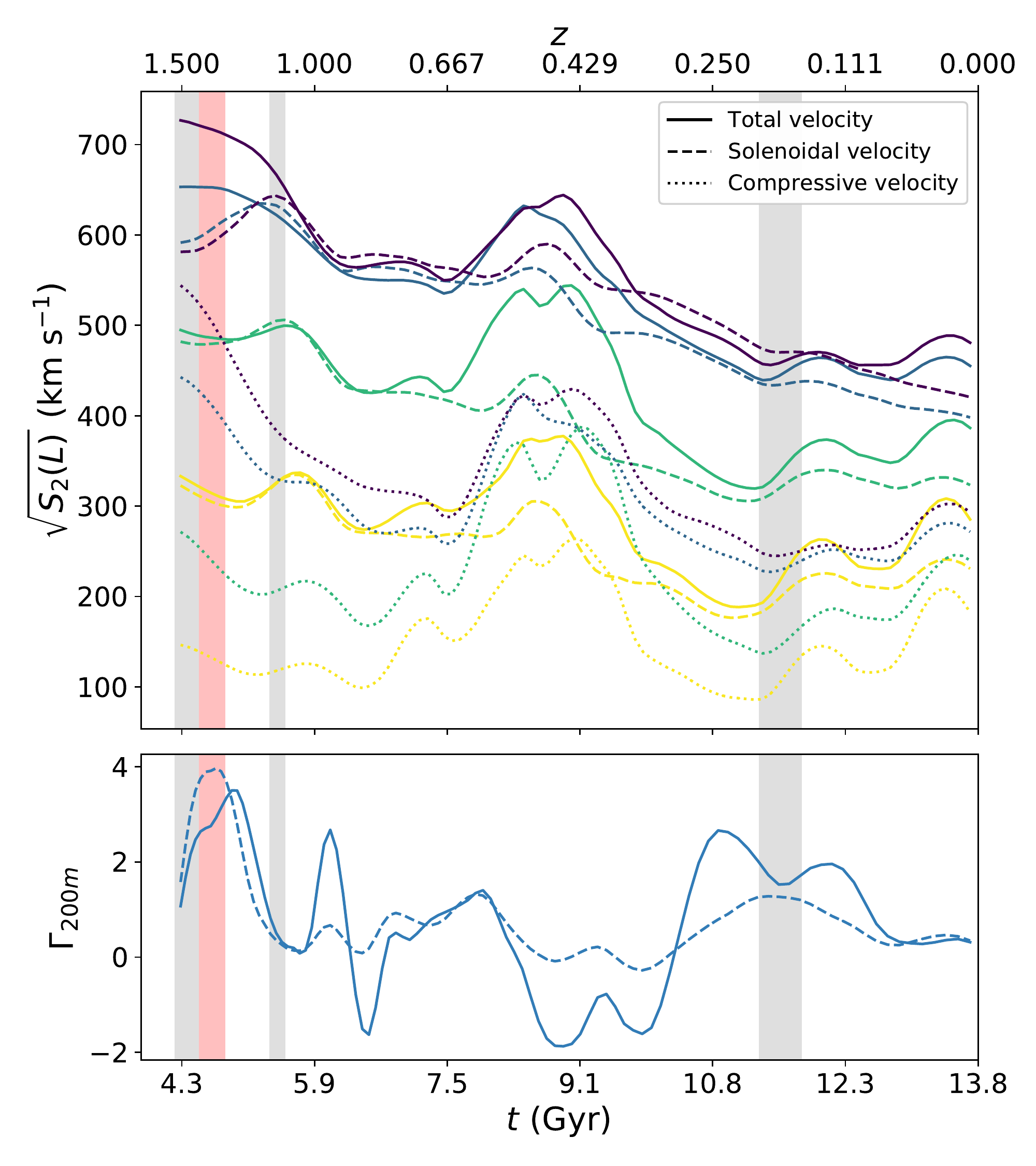}}
\caption{Top panels show the evolution, from $z\simeq1.5$ to $z\simeq 0$, of the velocity fluctuations on scales $L=0.1 R_\mathrm{vir}$ (yellow), $0.3 R_\mathrm{vir}$ (green), $R_\mathrm{vir}$ (blue) and $1.5 R_\mathrm{vir}$ (purple), for clusters CL01 (left) and CL02 (right). The lower panels show the baryonic (solid lines) and total (dashed lines) instantaneous MARs. The velocity fluctuation curves have been smoothed using a Savitzky-Golay filter of the same order and window length as the MARs, for both curves to have a similar level of (temporal) locality. The legends apply to both panels.}
\label{fig:evolution_velocity_fluctuations}
\end{figure*}

Figure \ref{fig:evolution_velocity_fluctuations} presents the joint analysis of velocity fluctuations on different scales, accretion rates and merging regimes. For the case of CL01 (left-hand panel), an important correlation between the evolution of velocity fluctuations on different scales and MARs is noticed. At high redshifts, while the cluster is fastly accreting gas from the mergers, velocity fluctuations are mantained at high values. In fact, the major merger at around $z \sim 0.9$ and the corresponding peak of the baryonic MAR at $z \sim 0.8$ injects a large amount of kinetic power on large scales, resulting in the enhancement of the corresponding $S_2(L)$ in the following $\sim \mathrm{Gyr}$ ($S_2(1.5 R_\mathrm{vir})$ peaks at $z \sim 0.67$, $1.1 \, \mathrm{Gyr}$ later than the MAR peak). The delay in the peaks at smaller scales directly reflects how the energy is cascading to smaller scales through fluid instabilities. For example, at $0.1 R_\mathrm{vir}$ scales, the total velocity fluctuation peaks at $z \sim 0.55$ ($800 \, \mathrm{Myr}$ after the peak at the largest scales). On the other hand, after the minor merger at $z \sim 0.15$, which does not have a severe impact on the MARs, the compressive velocity fluctuations are slightly enhanced at all scales (but not so the solenoidal ones). 

As for cluster CL02, the most remarkable feature is perhaps the increase of the velocity fluctuations at all scales during $0.7 \gtrsim z \gtrsim 0.4$. Incidentally, during this period the cluster presents negative baryonic MARs, while the DM halo mass remains fairly constant (and so the total MAR is close to zero). This effect could be due to, e.g., gas sloshing in the DM potential well, or due to a merger-accelerated shock scenario \citep{Zhang_2020}. Although we do not pursue a detailed explanation of the dynamical origin of this phenomenon, visual inspection of density and Mach number slices around the object, coupled with the fact that the increase in velocity fluctuations is driven by the compressive component of the velocity field, tend to suggest the latter as the most plausible.

The examples above depict a complex and varied phenomenology where, nevertheless, galaxy cluster mergers and accretion phenomena seem to dominate the evolution of the velocity fluctuations through cosmic time, acting thus as primary sources of turbulence in the ICM. In any case, these mechanisms could be punctuated, specially at the smallest scales and in the innermost regions of galaxy clusters, by sources of feedback such as SNe and/or AGN energy injection (the latter not been accounted for in this simulation).



\subsection{Enstrophy and helicity}
\label{s:global.vorticity_helicity}

We now consider the evolution of two quantities intimately related to solenoidal turbulence, which as we have seen is the dominant ICM turbulence component. These two quantities are computed from the pseudo-vector \textit{vorticity}, i.e., the curl of the velocity field, $\bm{\omega} = \nabla \times \vb{v}$. From this quantity, we consider the scalar \textit{enstrophy},

\begin{equation}
	\epsilon = \frac{1}{2} \bm{\omega}^2,
	\label{eq:enstrophy_def}
\end{equation}

\noindent which has been employed in many previous studies \citep{Porter_2015, Iapichino_2017, Vazza_2017, Wittor_2017, Valdarnini_2019} as a proxy of solenoidal turbulence; and the pseudo-scalar\footnote{In this paper, we refer to \textit{pseudo-scalar} quantities are those which change sign under parity, i.e., under the change of coordinates $\vb{x} \mapsto \vb{-x}$.} \textit{helicity} (e.g., \citealp{Moffatt_2014}),

\begin{equation}
	\mathcal{H} = \vb{v} \cdot \bm{\omega},
	\label{eq:vorticity_def}
\end{equation}

\noindent which is an interesting quantity from several points of view, as it will be discussed below, which however has not yet been applied, to our knowledge, to previous studies of cosmic flows. When focusing on a particular cluster, we shall compute $\mathcal{H}$ in its rest frame. The evolution of these quantities can be obtained, in the cosmological case, from the curl of the equation for the evolution of peculiar velocity in comoving coordinates. After some manipulation, this yields the following equation for the evolution of the (peculiar) vorticity pseudo-vector:

\begin{equation}
\begin{aligned}
	&\pdv{\bm{\omega}}{t} + \frac{1}{a}\left(\vb{v} \cdot \nabla \right) \bm{\omega} = \\&- H \bm{\omega} - \frac{1}{a}\bm{\omega} (\nabla \cdot \vb{v}) + \frac{1}{a}(\bm{\omega} \cdot \nabla) \vb{v} + \frac{1}{\rho^2 a} \nabla \rho \times \nabla P
	\label{eq:evolution_vorticity}
\end{aligned}
\end{equation} 

\noindent $P$ being the thermal pressure of the gas. That is to say, local vorticity changes due to advection, cosmic expansion, local fluid expansion, vortex stretching and baroclinicity (reading, from left to right, the terms in Eq. \ref{eq:evolution_vorticity}). From this equation, one can straightforwardly get the equation for the evolution of enstrophy in the comoving frame, since $\pdv{\epsilon}{t} = \bm{\omega} \pdv{\bm{\omega}}{t}$ and therefore

\begin{equation}
\begin{aligned}
	&\pdv{\epsilon}{t} + \frac{1}{a}\nabla \cdot \left(\epsilon \vb{v} \right) = \\&- 2 H \epsilon - \frac{1}{a} \epsilon (\nabla \cdot \vb{v}) + \frac{1}{a} \bm{\omega} (\bm{\omega} \cdot \nabla) \vb{v} + \frac{\bm{\omega}}{\rho^2 a} \left( \nabla \rho \times \nabla P \right).
	\label{eq:evolution_enstrophy}
\end{aligned}
\end{equation} 

Note this equation is equivalent to equation 3 in  \cite{Porter_2015}, but in this case in terms of the peculiar magnitudes of the fluid (removing the cosmological background), and without the magnetic and dissipative terms (since we do not include magnetic field nor we can explicitly evaluate the viscosity in our numerical scheme). The terms in Eq. \ref{eq:evolution_enstrophy} can be interpreted in a one-to-one correspondence to the ones in Eq. \ref{eq:evolution_vorticity}. Writing $\pdv{\epsilon}{t} = \sum_\mathrm{i} F_\mathrm{i}$, we have:

\begin{itemize}
	\item The advection term, $F_\mathrm{adv} = - \frac{1}{a}\nabla \cdot (\epsilon \vb{v})$, whose volume integral equates to the net enstrophy inflow.
	\item The cosmic expansion term, $F_\mathrm{cosm} = - 2 H \epsilon$, which dilutes the local enstrophy proportionally to the cosmic expansion rate, $H(t) \equiv {\dot a}/a$.
	\item The peculiar expansion/compression term, $F_\mathrm{pec} = - \frac{1}{a} \epsilon (\nabla \cdot \vb{v})$. Although we shall generally refer to this term as the `expansion' term, note it is positive (negative) wherever the fluid contracts, $\nabla \cdot \vb{v} < 0$ (expands, $\nabla \cdot \vb{v} > 0$). 
	\item The vortex stretching term, $F_\mathrm{vs} = \frac{1}{a} \bm{\omega} (\bm{\omega} \cdot \nabla) \vb{v}$, which appears when a gas element is accelerated in the direction of its vorticity (and therefore the vortex tube is stretched).
	\item The baroclinic term, $F_\mathrm{baroc} = \frac{\bm{\omega}}{\rho^2 a} \left( \nabla \rho \times \nabla P \right)$. Looking at the corresponding term in Eq. \ref{eq:evolution_vorticity}, this is the only term which can generate vorticity even if $\bm{\omega} = \vb{0}$.
\end{itemize}
 
Similarly, an equation for the evolution of helicity can be derived from Eq. \ref{eq:evolution_vorticity}, given that $\pdv{\mathcal{H}}{t} = \pdv{\vb{v}}{t} \bm{\omega} + \vb{v} \pdv{\bm{\omega}}{t}$:

\begin{equation}
\begin{aligned}
	&\pdv{\mathcal{H}}{t} + \frac{1}{a} \nabla \cdot \left(\mathcal{H} \vb{v}\right) = - 2 H \mathcal{H} + \frac{1}{2a} (\bm{\omega} \cdot \nabla) v^2 \\& + \frac{\vb{v}}{\rho^2 a} \cdot \left( \nabla \rho \times \nabla P \right) - \frac{1}{a} \nabla \cdot (\Phi \bm{\omega}) - \frac{1}{\rho a} \nabla \cdot (P \bm {\omega}).
	\label{eq:evolution_helicity}
\end{aligned}
\end{equation} 

Thus, the local helicity changes due to advection, cosmic expansion, vortex stretching and baroclinicity, as for the previous magnitudes (reading the terms from left to right; note, however, that there is no peculiar expansion/compression term). The last two terms represent the contribution of the acceleration of pre-existing vortices due to gravity and thermal pressure gradients. In the case of a barotropic fluid, i.e. $P = P(\rho)$, and in the limit where the term due to cosmology can be neglected (i.e., $H\mathcal{H}$ is much smaller than the rest of terms in Eq. \ref{eq:evolution_helicity}), the volume-integrated helicity is a conserved quantity (see, e.g., \citealp{Webb_2018}). Thus, the changes in helicity inside a cluster volume can be directly traced to either accretion (or, generally, mass flows) or baroclinicity.

\subsubsection{Evolution of the volume-averaged enstrophy and helicity}
\label{s:global.vorticity_helicity.evolution}

In order to study the driving mechanisms for the evolution of the volume-averaged helicity and enstrophy through our considered redshift interval, $1.5 \gtrsim z > 0$, we have computed the volume average of each of the source terms present in Eqs. \ref{eq:evolution_enstrophy} and \ref{eq:evolution_helicity}, respectively. In Appendix \ref{s:appendix.resolutionconvergence}, we discuss the convergence of the volume-averaged quantities. For each snapshot, we perform the integrals out to the virial radius (at the given snapshot; thus, time-dependent), since a pre-defined, constant comoving radius would imply accounting for a changing composition (in terms of phases), e.g., very non-virialized regions at earlier redshifts, when $R_\mathrm{vir}$ is smaller. In order to keep track of the magnitude of the \textit{pseudo-evolution} due to the time-dependent integration volume, we add an ad hoc pseudo-evolution source term, which is computed at each snapshot as the difference between the magnitudes average over the new and the previous integration domains. For conciseness, we only present the results for the most massive object, CL01. The results for CL02, although clearly different from the ones for the object we study here due to their different assembly histories, display similar trends. As for the integrated source terms, we have checked that our results do not depend strongly on resolution.

\begin{figure}
\centering
{\includegraphics[width=\linewidth]{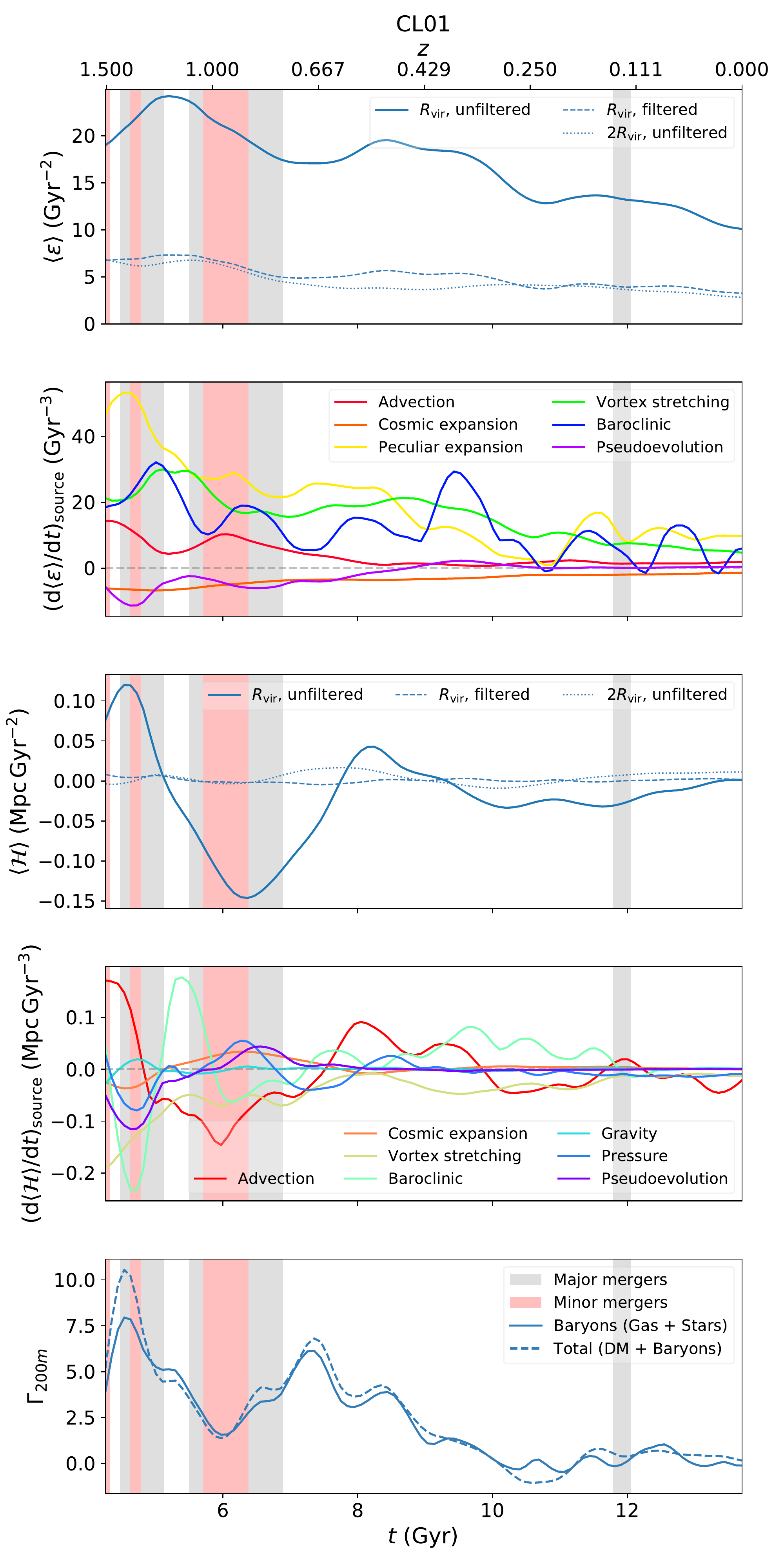}}
\caption{The first and third panels present the evolution of the volume-averaged enstrophy and helicity with cosmic time (or redshift, see legend in the upper panel), respectively. Solid and dashed lines present the unfiltered (total) and filtered (turbulent) quantities averaged over $R_\mathrm{vir}$, while dotted lines correspond to the unfiltered quantity inside the larger, $2R_\mathrm{vir}$, volume. The second and fourth panels quantify the importance of each of the source terms in Eqs. \ref{eq:evolution_enstrophy} (enstrophy) and \ref{eq:evolution_helicity} (helicity), respectively. The lower panel presents the baryonic (solid) and total (dashed) MARs, while background colours correspond to the merging regimes.}
\label{fig:evolution_enstrophy_helicity}
\end{figure}

Figure \ref{fig:evolution_enstrophy_helicity} presents the evolution of enstrophy (top panel), its source terms as defined in Eq. \ref{eq:evolution_enstrophy} (second panel), the evolution of helicity (third panel), its source terms as in Eq. \ref{eq:evolution_helicity} (fourth panel) and, for comparison, the MARs (bottom panel). The solid line in the upper panel represents the total (i.e., unfiltered) enstrophy averaged over the virial volume. This quantity undergoes strong evolution across the history of the cluster, intimately tied to the evolution of the accretion rates. Generally, the cluster-averaged enstrophy peaks after major mergers roughly 1 Gyr after the MARs do, thus displaying a delay effect similar to what we have seen in the evolution of the velocity fluctuations (Sec. \ref{s:global.strfunc}). At recent times, as the net accretion rate halts, the enstrophy falls to values around $10 \, \mathrm{Gyr^{-2}}$, which are consistent with the values \cite{Vazza_2017} obtain for a similar, slightly less massive cluster. Thus, stirring of the ICM due to mergers, and more generally accretion of gas, appears to be the primary source of solenoidal turbulence in CL01 \citep{Shi_2014, Nelson_2014, Miniati_2015}. Focusing on a larger comoving volume ($2R_\mathrm{vir}$ around the cluster, dotted lines in the same panel), the evolution of enstrophy is much milder, and enstrophy values themselves are lower. This is again due to, both, numerical and physical reasons. On the one hand, the larger cell sizes in clusters' outskirts suppress the energy cascade at larger scales, as vortices smaller than several times the local resolution length cannot be resolved. Although a handful of methods to overcome this inconvenience exist (either through subgrid modelling, e.g. \citealp{Schmidt_2016, Kretschmer_2020}; or through ad hoc refinement based on local vorticity, e.g. \citealp{IapichinoNiemeyer_2008}), the dynamical relevance of the possible vortices that we could miss in such low-density regions is reduced and, indeed, \cite{Vazza_2017} also find a reduced value of the enstrophy in their large volume, as lower density regions typically experience less intense dynamics. Dashed lines present the evolution of the turbulent enstrophy (i.e., the one computed from the small-scale [turbulent] velocity field, $\delta \bm{\omega} \equiv \nabla \times \delta \vb{v}$). While displaying a similar evolution to the corresponding quantity from the total velocity field, we find that the turbulent enstrophy contributes around $\sim 1/3$ to the total cluster-integrated enstrophy budget.

The source terms shown in the second panel correspond to the integrated, unfiltered enstrophy inside the virial volume (solid line in the upper panel). There are several prominent reasons for the sum of the source terms not to add up to the gradient of the enstrophy curve. First, even though we have not quantified it, numerical dissipation causes small vortices at the end of the turbulent cascade to fade, effectively acting as a sink term. Second, the source terms are computed on the sparsely saved simulation outputs (roughly, each $\sim 300 \, \mathrm{Myr}$), which does not allow to properly integrate them for the smallest scales (say, below $\sim 100 \mathrm{kpc}$, c.f. Fig. \ref{fig:timescales}). This same effect is, indeed, seen in, e.g., \cite{Porter_2015, Vazza_2017}, and also in \cite{Wittor_2017}, who investigate the Lagrangian evolution of enstrophy using passively advected tracer particles in post-processing\footnote{When comparing our results to \cite{Wittor_2017} note, however, that their results present much narrower and taller peaks, as they focus on different families of tracer particles mostly associated to gas clumps, while we focus on cluster-wide quantities.}. Nevertheless, the values at the sparsely saved snapshots can still be used to interpret the relative importance of the different mechanisms. 

The pseudo-evolution term (purple line) is generally small, compared to the rest of sources, implying that our definition of a Lagrangian region for quantifying enstrophy is not a bad choice and the changes in its value are dominated by cosmic flows and inner dynamics. Cosmic expansion, as well, is barely important when compared to the rest of mechanisms. Its effect is only marginally relevant ($\lesssim 10\%$) at higher redshifts, when $H(t)$ is larger. The primary mechanism of enhancement of cluster's enstrophy appears to be compression (yellow line; positive and thus $\left<\nabla \cdot \vb{v} \right> < 0$), which especially dominates during (major) merger epochs. The integral of the advection term (red line) corresponds to the net flux of enstrophy. This quantity is positive during the mergers, as the cluster is feeding from the already stirred ICM of the infalling clusters. As the accretion rates decrease, this term becomes irrelevant, slightly negative (as the gas is advected from lower density regions, where the enstrophy is lower). These terms, however, imply redistribution, but not properly speaking \textit{sources} of vorticity. Vortex stretching and baroclinicity are persistent and non-negligible during the whole cluster's history. While the former seems to be enhanced after the mergers, we do not see a clear trend for the later. We will come back to these sources in Sec. \ref{s:local.sources}, when examining their spatial distribution.
 
As for helicity, because of not being a positive-definite quantity, contrary to enstrophy, and being a pseudo-scalar, it should average to zero as long as the cluster is spherically symmetric (and, thus, invariant under reflections). Therefore, interpreting this quantity can be slightly more subtle. As we will see in Sec. \ref{s:local.enstrophy_helicity}, the role of helicity can be better disentangled locally. The global evolution of its volume-integrated value (third panel in Fig. \ref{fig:evolution_enstrophy_helicity}) reveals how mergers, again, induce strong variations in helicity. The fourth panel in Fig. \ref{fig:evolution_enstrophy_helicity} shows the helicity source terms evolution. Advection dominates the helicity variations during the mergers period, as the cluster accretes the ICM of the merged clusters which may already have developed some degree of helicity. Once discussed this more `geometric' term, baroclinicity and vortex stretching remain as the primary sources of helicity generation, while the acceleration of pre-existing vortices due to gravity and pressure gradients is slightly smaller in magnitude. Nevertheless, the evolution is fairly more complex than what we have seen for enstrophy. As covered for the enstrophy evolution, cosmic expansion and pseudo-evolution have a generally small contribution. Note how the helicity corresponding to the turbulent velocity field averages to zero over the cluster's volume. That behaviour is expected since, once the bulk flows (e.g., almost laminar infall of gas in the outskirts) have been removed, the residual velocity field is much more isotropic.

\subsubsection{Phase maps}
\label{s:global.vorticity_helicity.phase_maps}
\begin{figure*}
\centering
{\includegraphics[width=0.33\linewidth]{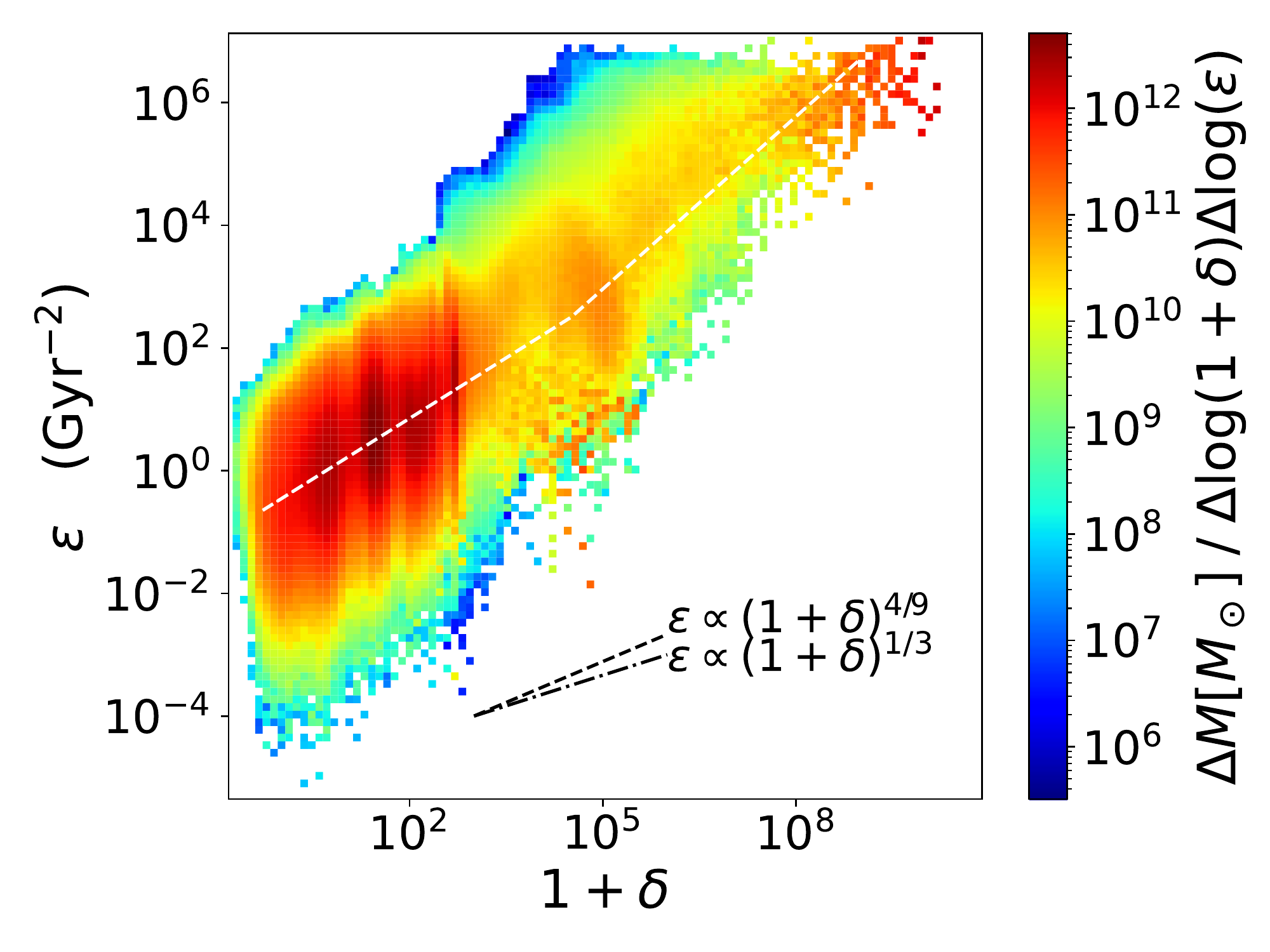}}~
{\includegraphics[width=0.33\linewidth]{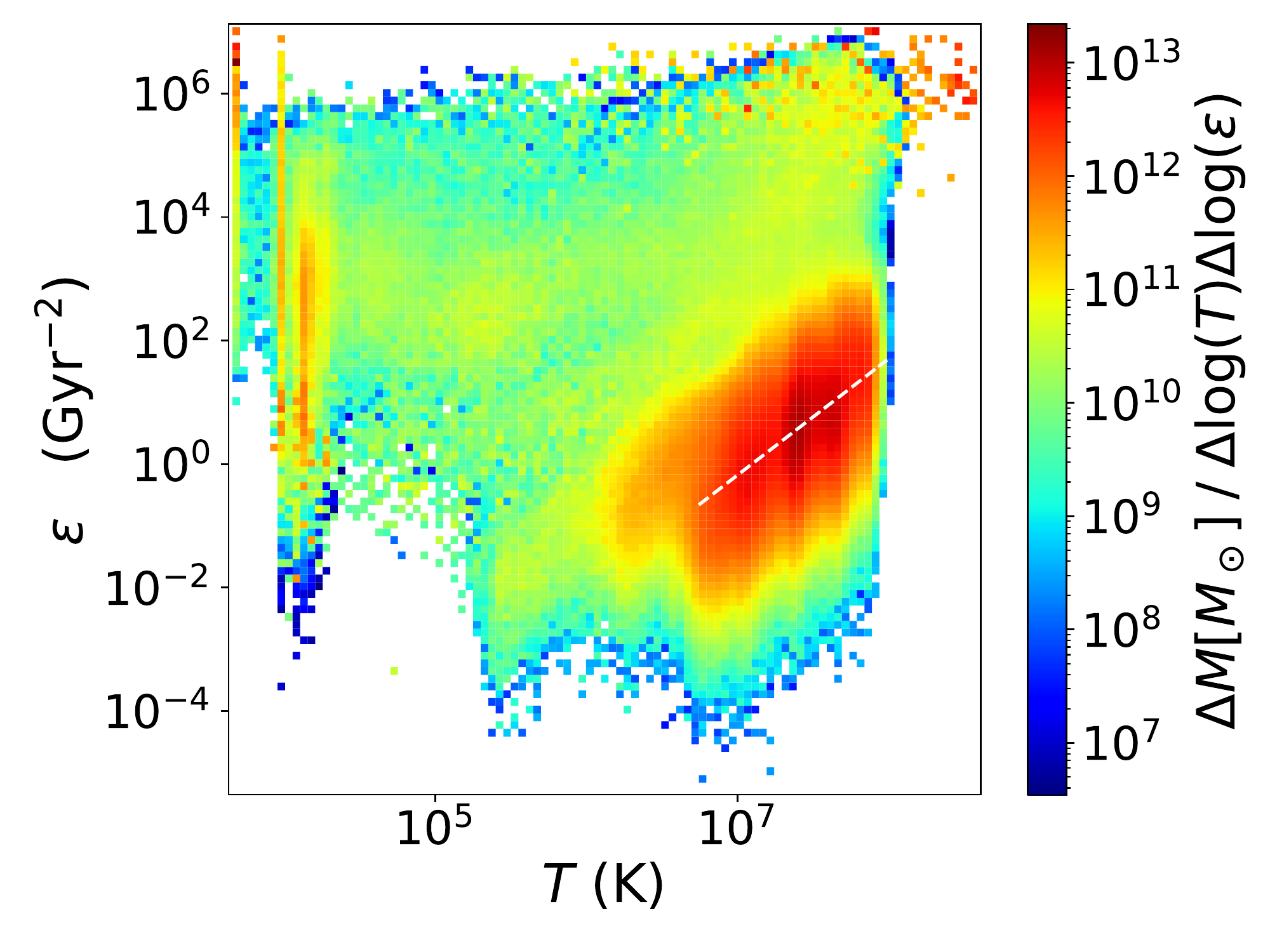}}~
{\includegraphics[width=0.33\linewidth]{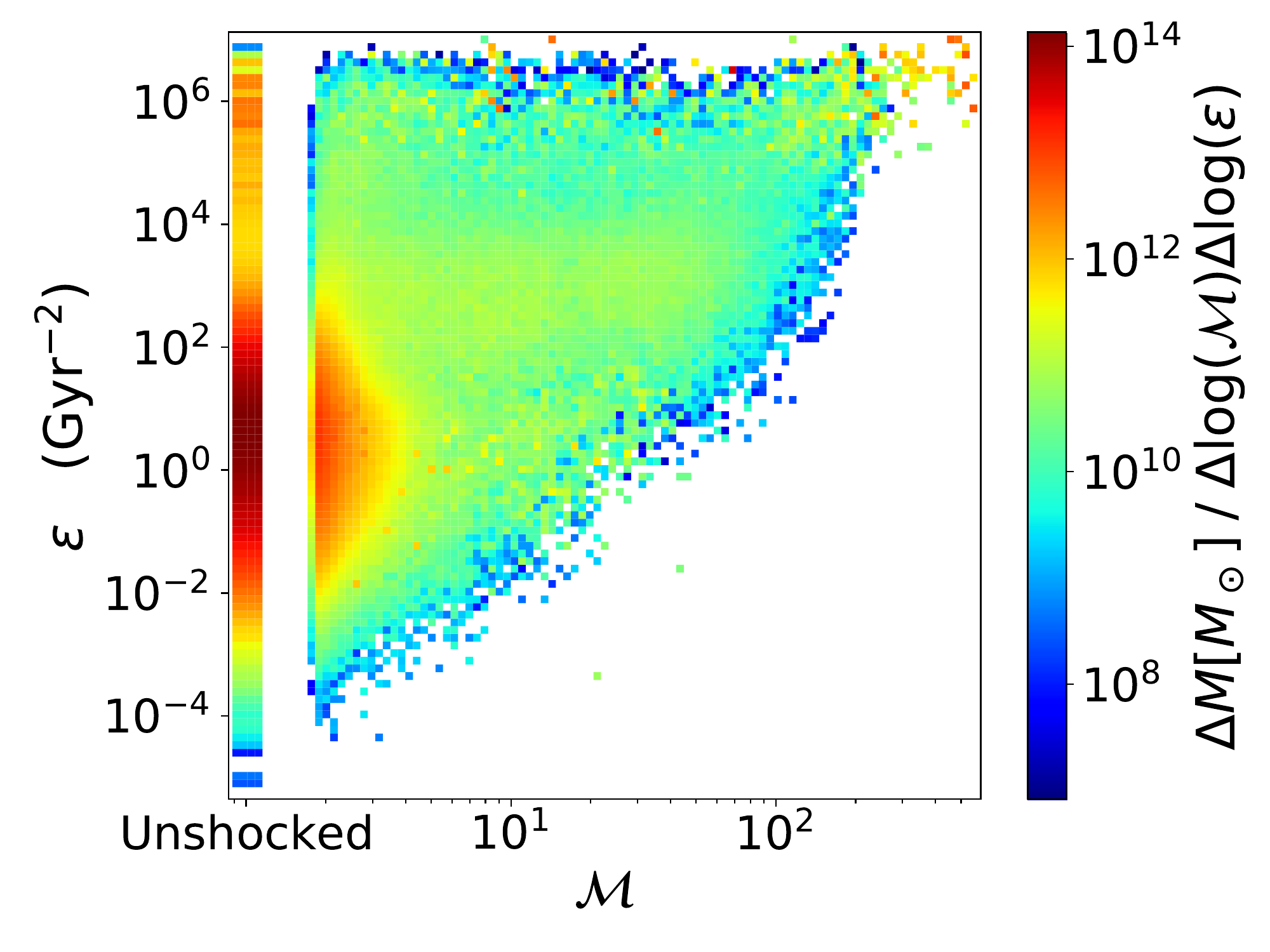}} 
{\includegraphics[width=0.33\linewidth]{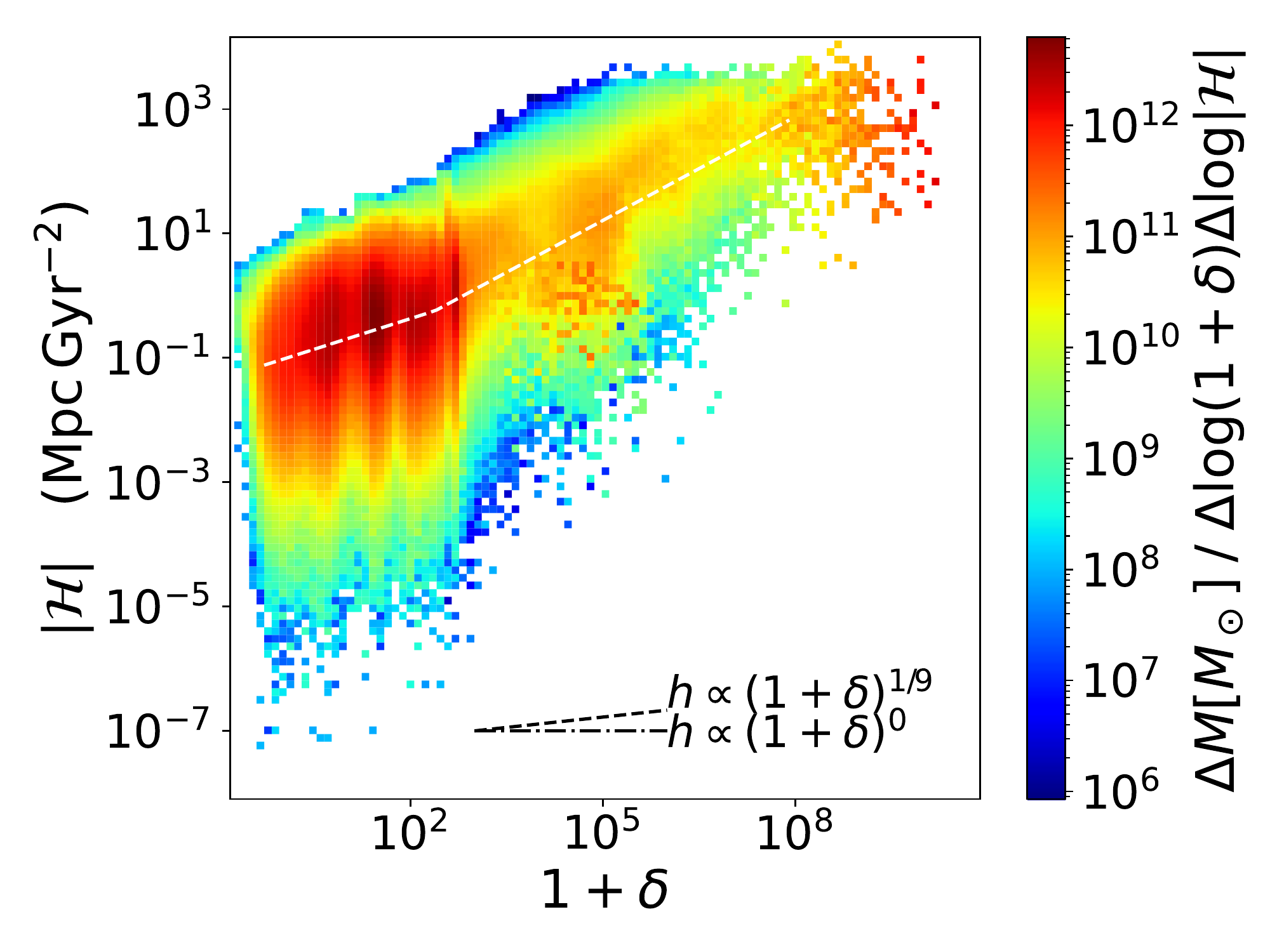}}~
{\includegraphics[width=0.33\linewidth]{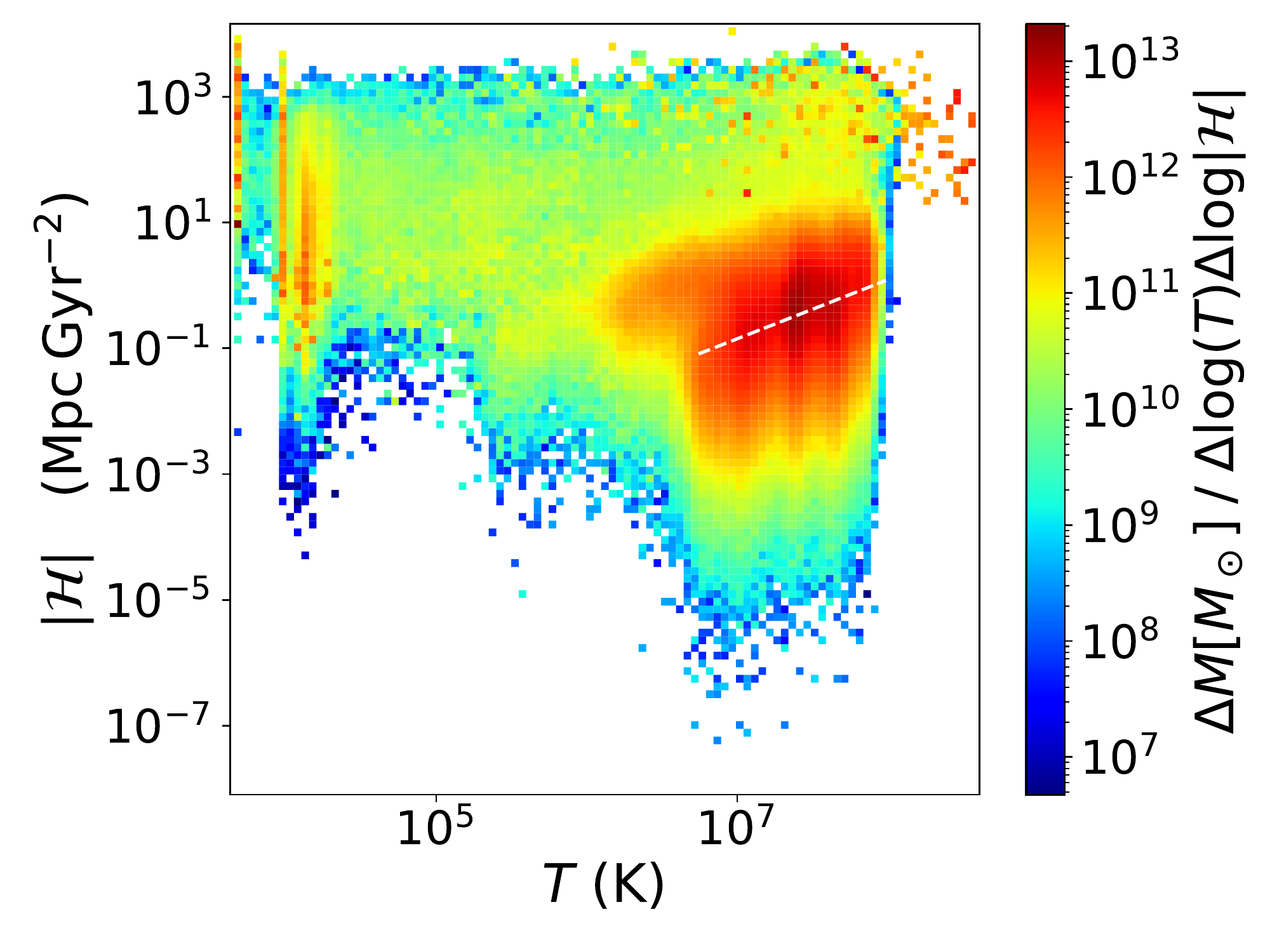}}~
{\includegraphics[width=0.33\linewidth]{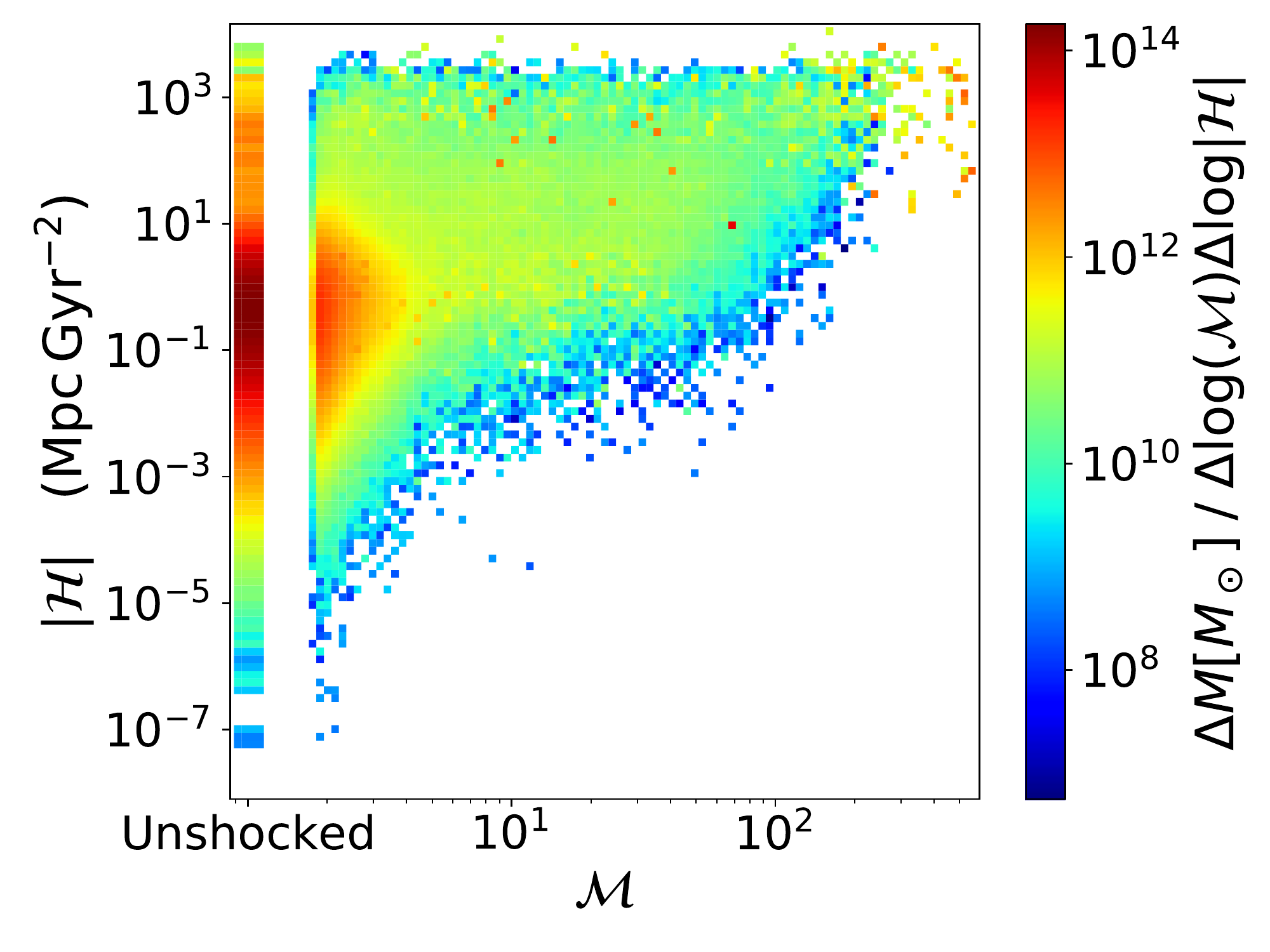}} 
\caption{Phase-space density maps, for CL01 at $z \simeq 0$, of enstrophy (top row, $\epsilon$) and helicity magnitude (bottom row, $|\mathcal{H}|$) vs. overdensity (left-hand column, $1+\delta \equiv \rho/\rho_B$), temperature (central column, $T$) and Mach number (right-hand column, $\mathcal{M}$). All fluid elements inside $2R_\mathrm{vir}$ are considered for producing the phase map. The color encodes the phase-space density (the hotter the color is, the larger the phase-space density). The black lines in the left column represent the expected slope if there is no physical correlation between the variables, and the only effect is the resolution one, assuming a Kolmogorov (dashed) and a Burgers (dash-dotted) spectrum. The white lines correspond to broken power-law fits (left-hand column) and simple power-law fits (middle column). The Mach number axis scale is cut at $\mathcal{M} \simeq 1.3$ and cells below this threshold are regarded as `unshocked'.}
\label{fig:phase_maps}
\end{figure*}

Finally, we explore the phase-space distribution of enstrophy and helicity against several thermodynamical quantities of the gas (namely, density and temperature; from which the scaling with other quantities, e.g. pressure or entropy, can be directly derived), as well as Mach number to elucidate their relation to shocks. We shall as well restrict the presentation of our results to the cluster CL01 at the most recent snapshot, at redshift $z \simeq 0$. For this aim, we present in Figure \ref{fig:phase_maps} the phase-space density maps for the aforementioned variables. In all cases, we have considered all the volume elements lying inside $2 R_\mathrm{vir}$, at the best resolution available at each point.
 
The left-hand panels in Fig. \ref{fig:phase_maps} represent the enstrophy-overdensity (upper panel) and helicity-overdensity (lower panel) phase-space densities. The general trend of both maps shows a tendency for denser gas to be more vortical (higher enstrophy) and helical (higher helicity). A relevant question in this scope, however, is whether this trend is physical or only due to a resolution effect (i.e., denser regions get resolved by smaller cell sizes and smaller vortices can develop, thus enhancing the vorticity magnitude). This effect can be ruled out by a simple scaling argument. If the velocity field scales according to a \cite{Kolmogorov_1941} spectrum, then one expects the vorticity to depend on the scale $L$ as $\omega_L \sim L^{-2/3}$. On the other hand, the pseudo-Lagrangian refinement scheme provides $(1+\delta) \sim L^{-3}$. Therefore, the dependence of vorticity on density due to pure resolution effects ought to be $\omega \sim (1+\delta)^{2/9}$. Bringing this result to the enstrophy and helicity definitions, the expected scaling is $\epsilon \sim (1+\delta)^{4/9}$ and $\mathcal{H} \sim (1+\delta)^{1/9}$. If we assumed a steeper, \cite{Burgers_1939} spectrum, in which $\omega_L \sim L^{-1/2}$, then we shall expect shallower dependencies, $\epsilon \sim (1+\delta)^{1/3}$ and $\mathcal{H} \sim (1+\delta)^{0}$. As ICM turbulence typically lies in between these regimes (e.g., \citealp{Miniati_2014}, \citealp{Vazza_2017}), if the dependency on density could be explained by resolution alone, then $\epsilon$ and $\mathcal{H}$ are expected to scale with logarithmic slopes between these two predictions.

The bulk of ICM's mass sits in the overdensity range $1+\delta \lesssim 10^3-10^4$. In this range, the logarithmic slope of the $\epsilon-(1+\delta)$ phase-space density map is much shallower than through the whole overdensity range, but still above the expectation by the scaling reasoning due to resolution. We have fitted the mass-weighted mean $\epsilon-(1+\delta)$ relation to a broken power law, which yields a logarithmic slope $0.658\pm 0.038$ for $1+\delta \lesssim 10^4$ (see white, dashed lines in the figure), thus pointing at a genuinely physical effect beyond resolution. A similar pattern is observed for helicity magnitude, whose observed slope for the bulk ICM ($0.329\pm0.059$) is inconsistent with the resolution effect alone. This result is not surprising, since denser gas (usually located towards smaller clustercentric radii, c.f. Fig. \ref{fig:structure_functions_phases}) varies on smaller spatial and temporal scales.

When relating our two measures of solenoidal turbulence, $\epsilon$ and $|\mathcal{H}|$, to temperature (central column in Fig. \ref{fig:phase_maps}), the resolution effect becomes less relevant. This is mainly due to the fact that the radial dependence of temperatures on the ICM is much weaker than that of densities (see, e.g., the profiles in \citealp{Planelles_2009}). There is a strong tendency for most of the cluster gas mass to be increasingly vortical for higher temperatures. In the region where most of the cluster's mass is concentrated, enstrophy follows a steep power law, with slope $1.884\pm0.065$. A similar behaviour is seen in the $|\mathcal{H}|-T$ map, with roughly half the slope (as $|\mathcal{H}| \propto \omega$ and $\epsilon \propto \omega^2$). However, if the whole phase-density map is considered, the low-helicity end presents a drop at $T \gtrsim 10^6 \mathrm{\,K}$, which is not clearly reflected in enstrophy. This shows that helical motions are partially suppressed for hotter gas. This effect can be much more easily interpreted in terms of the spatial distribution (Sec. \ref{s:local.enstrophy_helicity}), since hotter gas tends to reside, preferentially, in more central cluster regions (i.e., inside $R_\mathrm{vir}$).

The right column of Fig. \ref{fig:phase_maps} corresponds to the phase-space density of enstrophy and helicity magnitude with respect to the Mach number of shocked cells. Unshocked (and weakly shocked, $\mathcal{M} \lesssim 5$) cells have a mass-weighted enstrophy distribution which peaks around $\epsilon \sim 1-10 \, \mathrm{Gyr^{-2}}$, in agreement with the values displayed in Fig. \ref{fig:evolution_enstrophy_helicity} at recent redshifts. However, while most of the cluster mass resides in unshocked and weakly shocked regions, as strong shocks are only dominant in low-density environments (as the accretion shocks in the outskirts), we find these regions to have high levels of enstrophy, in spite of being resolved with lower spatial resolution. This would, indeed, suggest that we are capturing turbulent motions in the cluster outskirts, despite not being implementing any ad hoc refinement scheme (c.f., \citealp{Vazza_2009}, \citealp{Iapichino_2017}). The fundamental difference, which allows us to recover a high level of vorticity even in the outskirts, is the base grid resolution (which, in turn, determines the ``mass resolution'' of the pseudo-Lagrangian AMR approach). While the base grid cells in \cite{Iapichino_2017} are $2 \, \mathrm{h^{-1}Mpc}$ on each side, ours are $\sim 310 \, \mathrm{kpc}$. Therefore, this scheme is able to resolve with considerably higher resolution these outer regions without the need of additional refinement criteria. 

\begin{figure*}
\centering
{\includegraphics[width=0.33\linewidth]{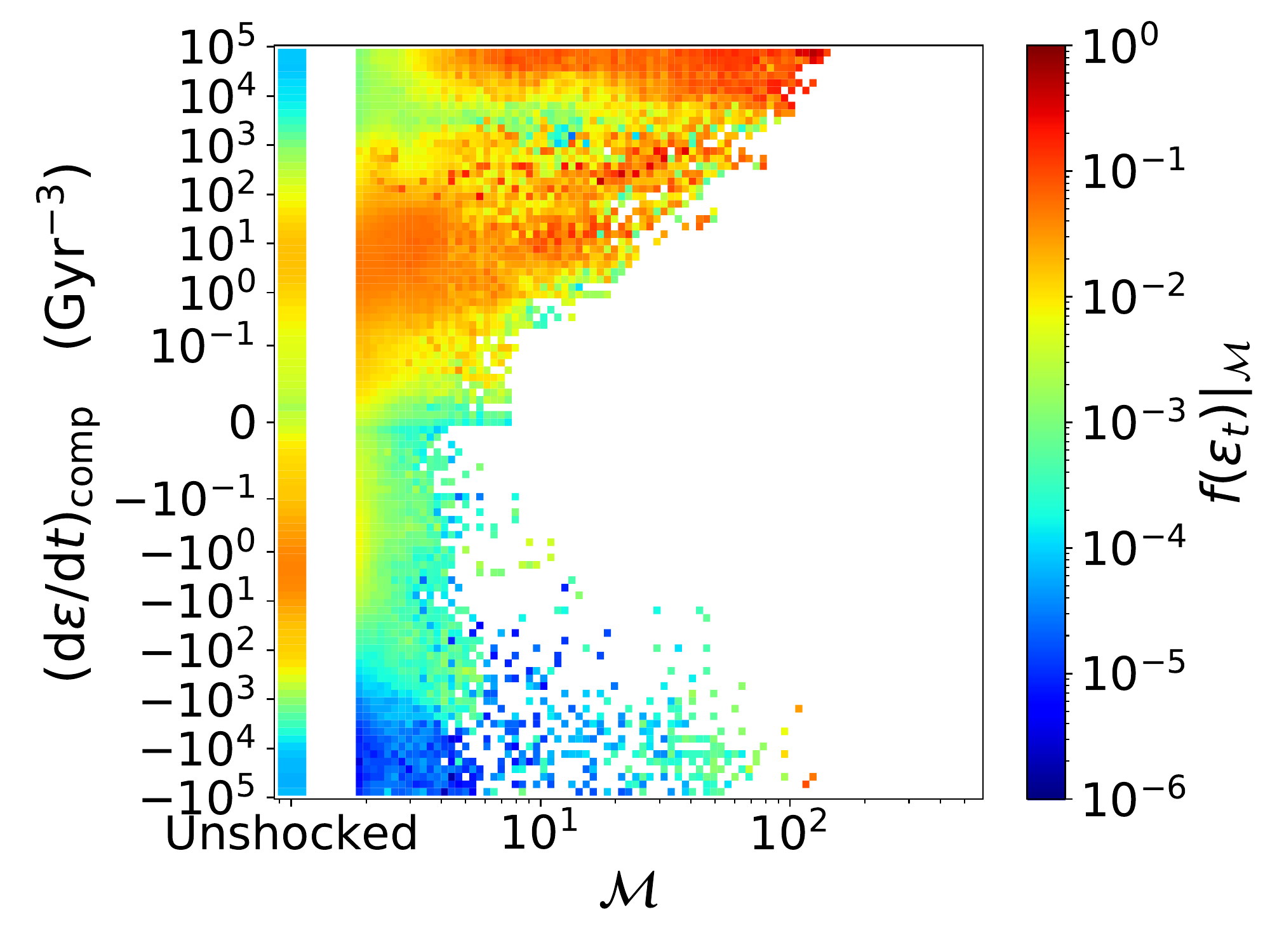}}~
{\includegraphics[width=0.33\linewidth]{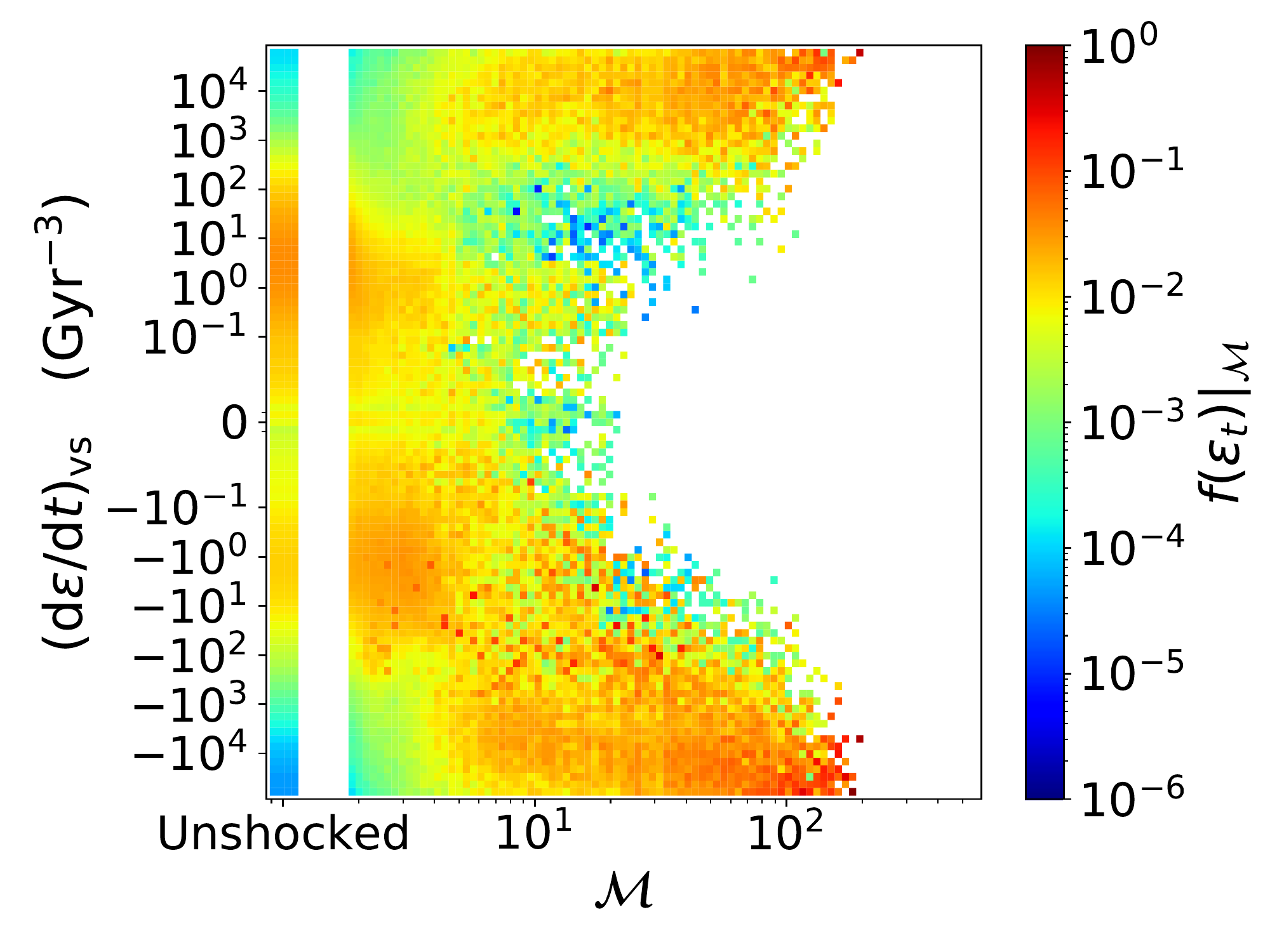}}~
{\includegraphics[width=0.33\linewidth]{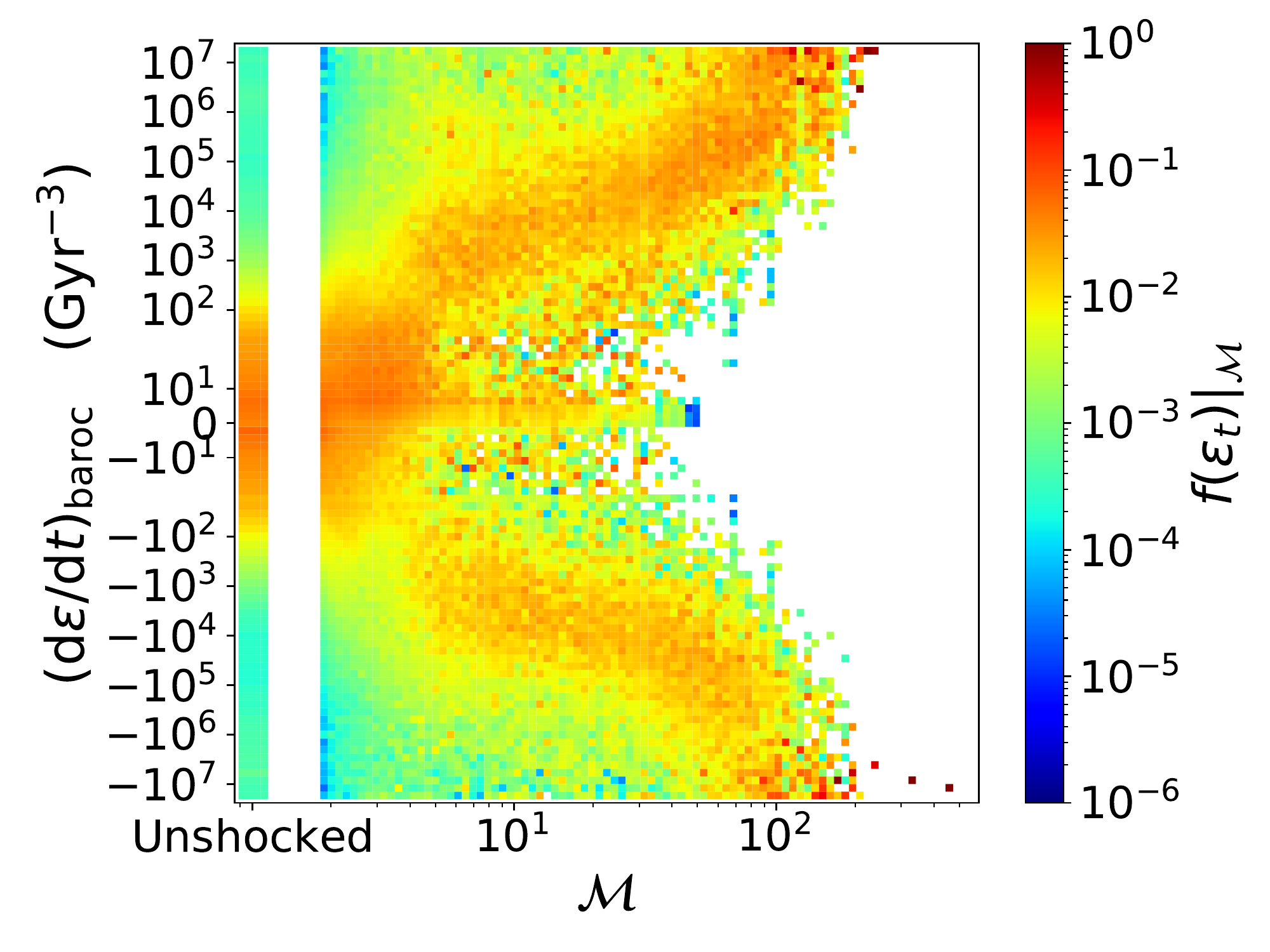}} 
\caption{Conditional distributions, for CL01 at $z \simeq 0$, of the compressive (left-hand panel), vortex stretching (center panel) and baroclinic (right-hand panel) enstrophy source terms strength as a function of the shock Mach number. That is, each column expresses the probability distribution of the values of the source terms, for all cells within $2R_\mathrm{vir}$ within a given $\mathcal{M}$ range. Mach number axis scale is cut at $\mathcal{M} \simeq 1.3$ and cells below this threshold are regarded as `unshocked'.}
\label{fig:phase_maps_sources}
\end{figure*}

In order to elucidate and disentangle the role of the main enstrophy source terms, we present in Figure \ref{fig:phase_maps_sources} the conditional distributions of the source terms as a function of the Mach number, which we have denoted $f(\epsilon_t) |_\mathcal{M}$ in the figures. We have focused on peculiar expansion/compression (left-hand panel), vortex stretching (middle panel) and baroclinicity (right-hand panel), which, as it has been seen in Sec. \ref{s:global.evolution.accretion}, are the dominant sources of cluster-wide enstrophy variation. We show conditional distributions, instead of phase maps as done above, since it better serves our purpose of analysing the trends of the enstrophy production/dissipation rates as a function of the shock's Mach number.

The compressive component is perhaps the simplest one to analyse. While unshocked cells have a relatively symmetric distribution of $\left(\pdv{\epsilon}{t}\right)_\mathrm{comp}$, with a subtle preference towards negative values (expansion), shocked cells are, unsurprisingly, dominated by positive values of the source term, since these cells have $\nabla \cdot \vb{v} < 0$. The extremely high values of $\left(\pdv{\epsilon}{t}\right)_\mathrm{comp}$ for strong shocks ($\mathcal{M} \gtrsim 10$), which are mostly located in the outskirts of the cluster, highlight the thermodynamically irreversible compression happening at the external, accretion shocks as a primary mechanism for enstrophy generation.

The distribution of values of the vortex stretching enstrophy source mechanism presents a clear distinction between unshocked and weakly shocked gas ($\mathcal{M} \lesssim 5$) and strong shocks ($\mathcal{M} \gtrsim 5$), and is fairly symmetrical under a change of sign, only with a subtle tendency towards more negative values in strong shocks. As it will be seen in Sec. \ref{s:local.sources}, this term is typically negative inside of the external shocks. 

The baroclinic term also presents a more or less symmetric distribution with respect to its sign, being slightly more positive (for all $\mathcal{M}$). Its magnitude clearly increases with the shock Mach number, although in this case the trend is much more continuous than for the vortex stretching term. Especially in the strongest shocks, baroclinicity is very efficient in inducing changes in vorticity, with a net effect of increasing the enstrophy. This is expected since, for strong enough shocks, the Rankine-Hugoniot jump conditions (see, e.g., \citealp{Landau_1987}) imply a mild density gradient but a steep pressure gradient. Therefore, the pressure gradient will be strongly aligned with the shock normal direction, while the density gradient can more easily have a non-negligible component tangential to the shock.


\section{Local description of turbulence}
\label{s:local}
Even though global statistics are informative of the general state of the cluster and are certainly useful when extending these analysis to large samples of clusters in order to attain statistically meaningful conclusions, some features associated with turbulence are much better captured by looking at their local distribution. In this section, we especially focus on solenoidal turbulence, which as we have seen is the dominant component. In particular, we analyse the distribution and relations amongst (filtered and unfiltered) enstrophy and helicity (Sec. \ref{s:local.enstrophy_helicity}), and the enstrophy source terms (Sec. \ref{s:local.sources}).

\subsection{Distribution of enstrophy and helicity}
\label{s:local.enstrophy_helicity}
\begin{figure*}
\centering
{\includegraphics[height=0.28\linewidth]{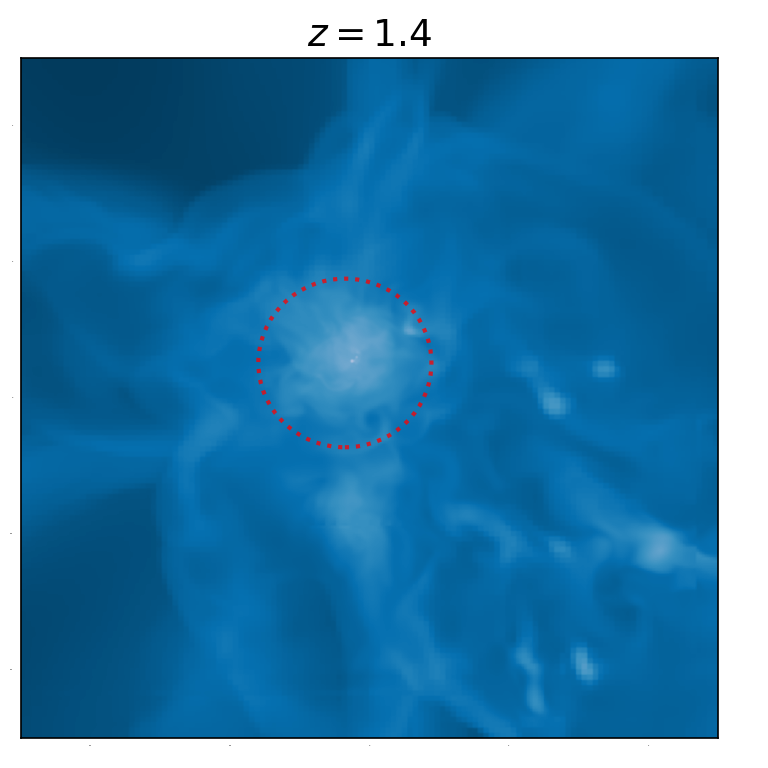}}~
{\includegraphics[height=0.28\linewidth]{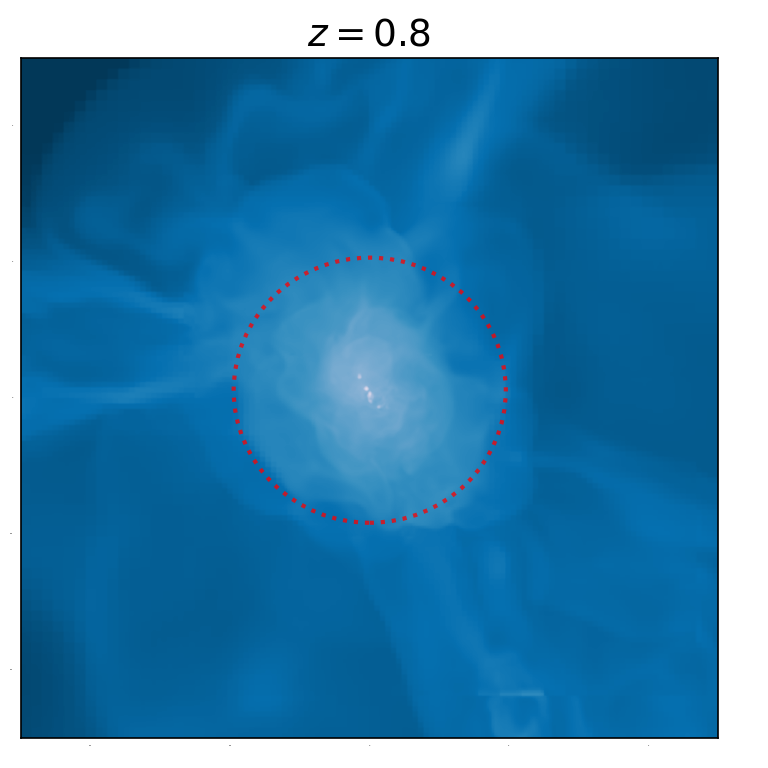}}~
{\includegraphics[height=0.28\linewidth]{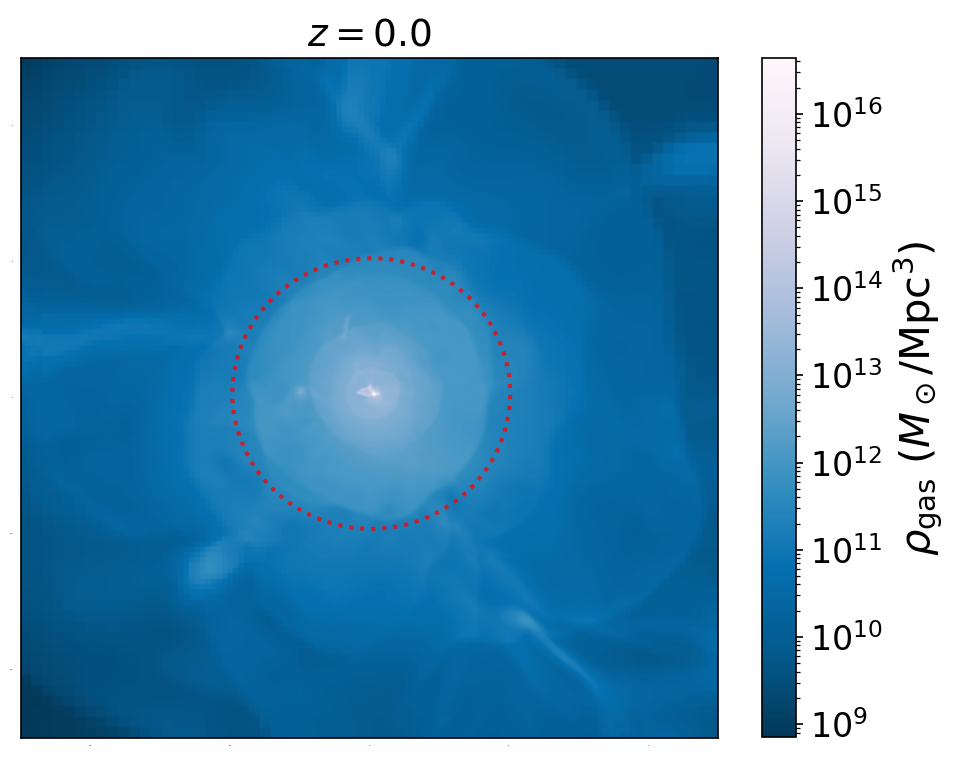}} 
{\includegraphics[height=0.28\linewidth]{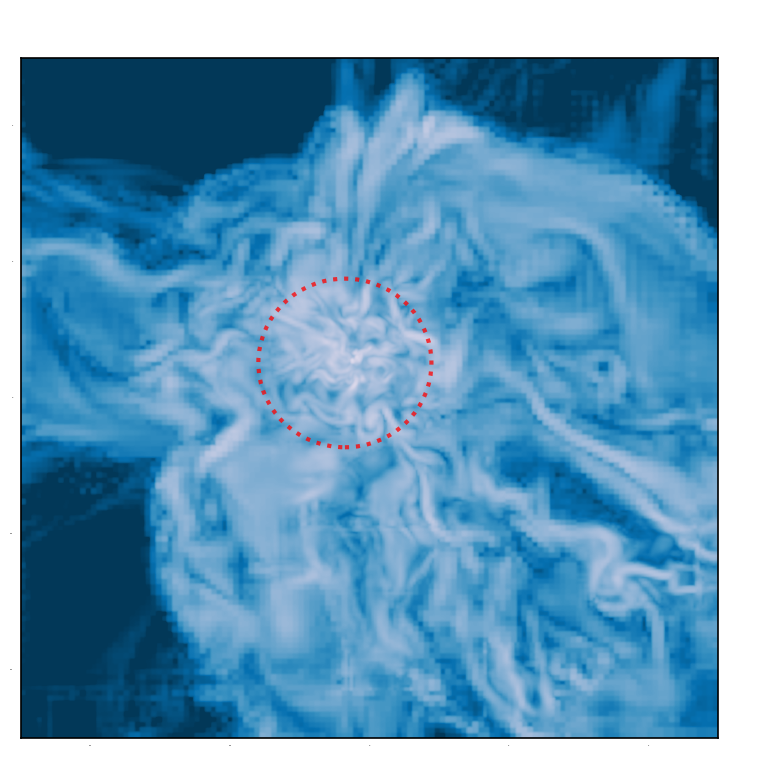}}~
{\includegraphics[height=0.28\linewidth]{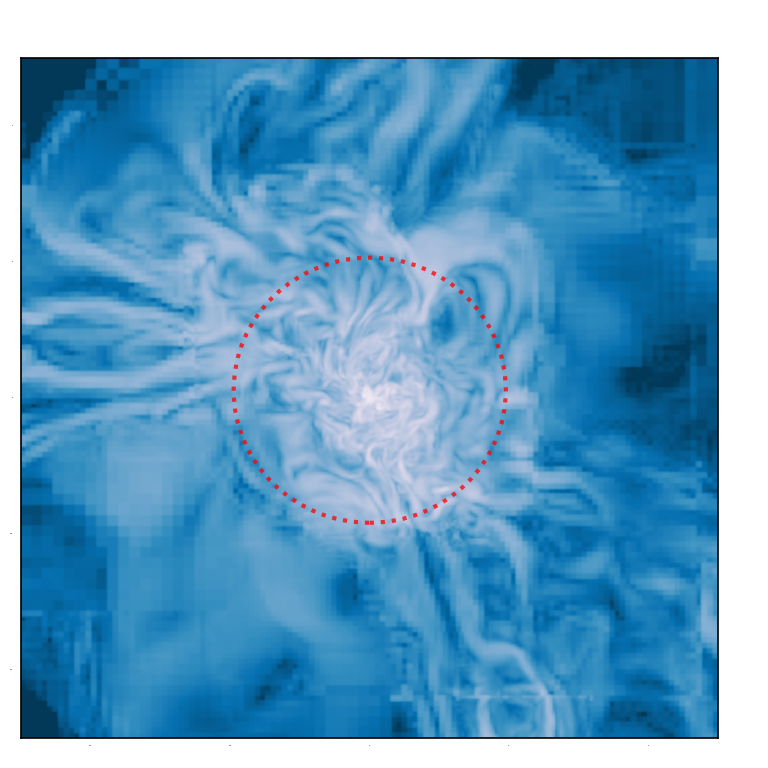}}~
{\includegraphics[height=0.28\linewidth]{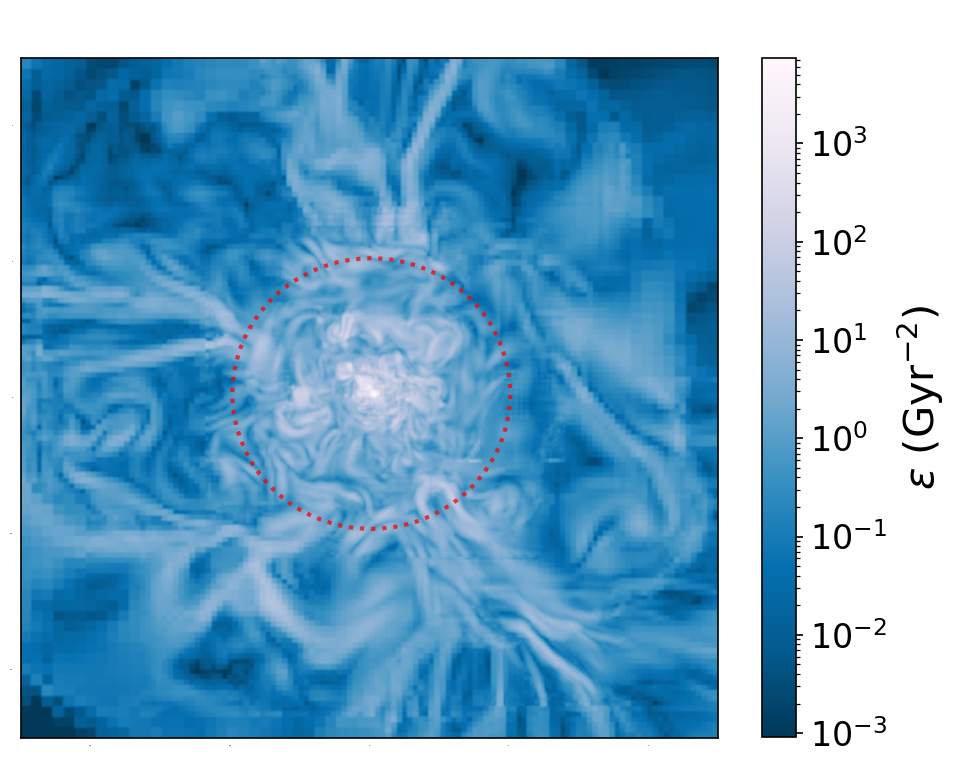}} 
{\includegraphics[height=0.28\linewidth]{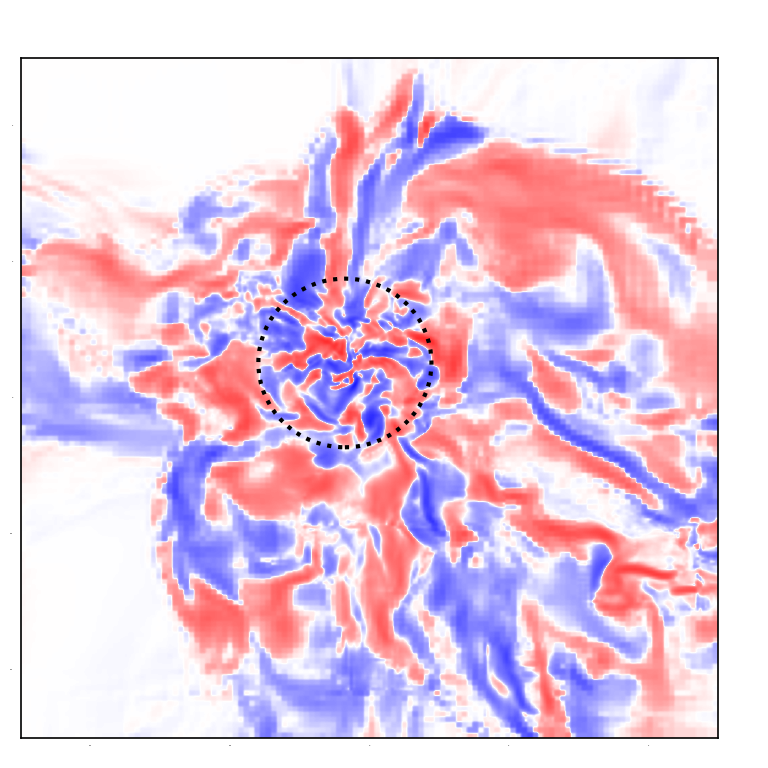}}~
{\includegraphics[height=0.28\linewidth]{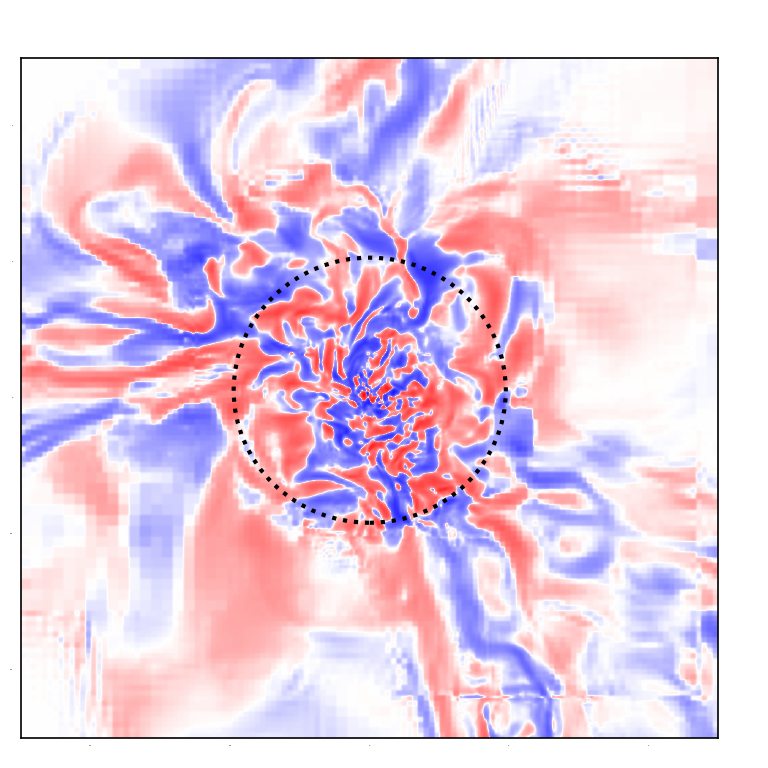}}~
{\includegraphics[height=0.28\linewidth]{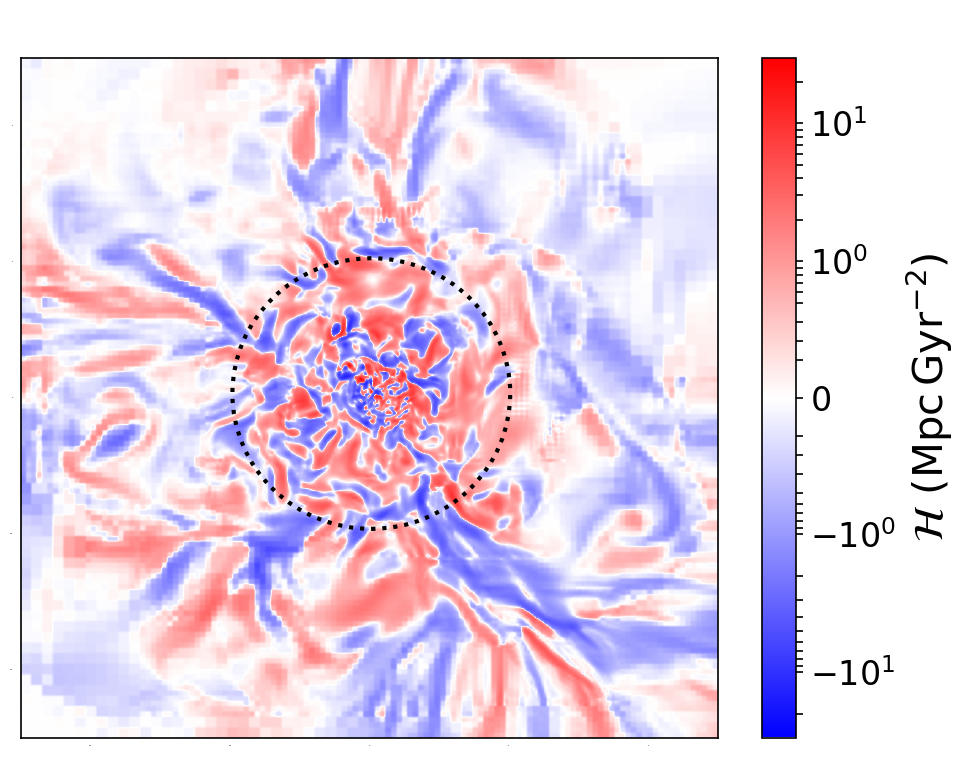}} 
\caption{Maps of gas density (top row), enstrophy (middle row) and helicity (bottom row) of a thin ($\sim 150 \, \mathrm{kpc}$ comoving) slice through cluster CL01 at three different redshifts: $z \simeq 1.45$ (left-hand panels) and $z \simeq 0.80$ (middle panels), while the cluster is undergoing strong accretion associated with mergers, and $z \simeq 0$, while the cluster is relaxed with negligible accretion rates. Each panel represents the same comoving region, $10 \, \mathrm{Mpc}$ (comoving) on each side.}
\label{fig:maps}
\end{figure*}

In Sec. \ref{s:global.vorticity_helicity}, it has been seen that denser and hotter gas tends to be more vortical and helical. To complement this, we now study the spatial distribution of these two magnitudes inside and around the cluster. In Figure \ref{fig:maps}, we show thin slices through the centre of the cluster of gas density (upper panels), enstrophy (middle panels) and helicity (lower panels); and for three redshifts: at $z\simeq 1.45$, $0.80$ (both just after major mergers) and $z\simeq 0$ (during a smooth accretion phase), respectively, in the columns left to right. In all plots, we indicate the virial volume with a dotted circle. 
 
At high redshift, the cluster, which has a noticeably smaller virial radius ($1.36 \, \mathrm{Mpc}$) and around a quarter of its total mass at $z=0$, is undergoing multiple mergers, resulting in an aspherical shape. The strongest external shocks, located at more than $2R_\mathrm{vir}$ in most directions, are clearly visible as density discontinuities, and so are the filaments, which are a dominant contribution to the accretion flows. The main merger has been produced by the infall of a structure from the right of the represented slice, where the cluster atmosphere is more extended. Complex gas structures appear in this region due to the stripping of the merged cluster's gas. A large part of this gas, which has ended up lying beyond $R_\mathrm{vir}$, will be reaccreted in the subsequent $\sim \mathrm{Gyr}$, mantaining the MARs high \citep{Valles_2020}.

At this stage, enstrophy is clearly bounded by the external shocks, especially in those regions where there have been no mergers by (left and top borders of the shown slices). This suggests that the baroclinic mechanism in the accretion shocks drives the first generation of vorticity to the hitherto pristine gas (since it is the only term in Eq. \ref{eq:evolution_vorticity} whose magnitude is not proportional to the vorticity magnitude), closely followed by strong compression which further enhances its magnitude. Filaments also present high levels of enstrophy, which most likely is the main contribution to the advection term. However, correctly accounting for and interpreting the contribution of filamentary accretion of gas may be complex and we leave it for future work.

Helicity naturally follows a similar pattern. Being defined as a pseudoscalar quantity, it presents ubiquitous changes of sign (especially around velocity discontinuities), but it is however an interesting quantity to explore the infall of matter to the cluster. Note that the filaments (better discernible in the density map, towards the left and top borders of the slice) are considerably helical: while enstrophy spans several orders of magnitude from the core to the filaments, filaments have only $\sim 1/10$ helicity magnitudes, compared to these of the core. We will explore this effect in more detail below, when showing the normalized radial profiles. 

After the second relevant major merger sequence has taken place, at $z\simeq 0.8$ (central column), the cluster has considerably increased its virial radius after reaccreting the dark matter particles and baryons which ended up beyond the spherical overdensity boundary, and the shape has become slightly more spherical. The main merged substructure has infallen from the bottom-right direction in the represented slice. In that direction, there is a clear gradient of enstrophy magnitude: while the gas near to the resulting cluster's virial boundary has considerably high levels of enstrophy with a complex spatial distribution, enstrophy is greatly reduced behind the shock. This can be interpreted as an evidence of the predominance of the compressive source term during mergers: enstrophy is being generated in the head-on collision of the two ICMs, primarily by the weak shocks arising during the merger. Besides the merger region, gas is also being intensely accreted through most of the cluster boundary. Note, for example, in the top-left of the slice, how the magnitude of helicity is enhanced towards the outskirts, where gas is infalling more laminarly. The complex patterns in helicity inside the virial volume are, at least partially, due to the presence of internal shocks and shear motions, which produce the frequent sign changes. 

At $z \simeq 0$, no important mergers have occurred recently and the ICM has relaxed to a nearly spherical shape, with extended accretion shock radii as well. A series of inner shocks, propagating outwards, are reflexed as density discontinuities (see also the compressive and solenoidal velocity magnitude maps in \citealp{Valles_2021_cpc}). Solenoidal motions, showing eddies on a wide range of scales, dominate in the volume bounded by the outermost shocks; not only in the central regions of the cluster, but also near the shocks and around filaments. Helicity is especially enhanced in these filaments, while its magnitude is less relevant elsewhere (due to the absence of bulk flows in the collapsed regions).  

\begin{figure}
\centering
{\includegraphics[width=\linewidth]{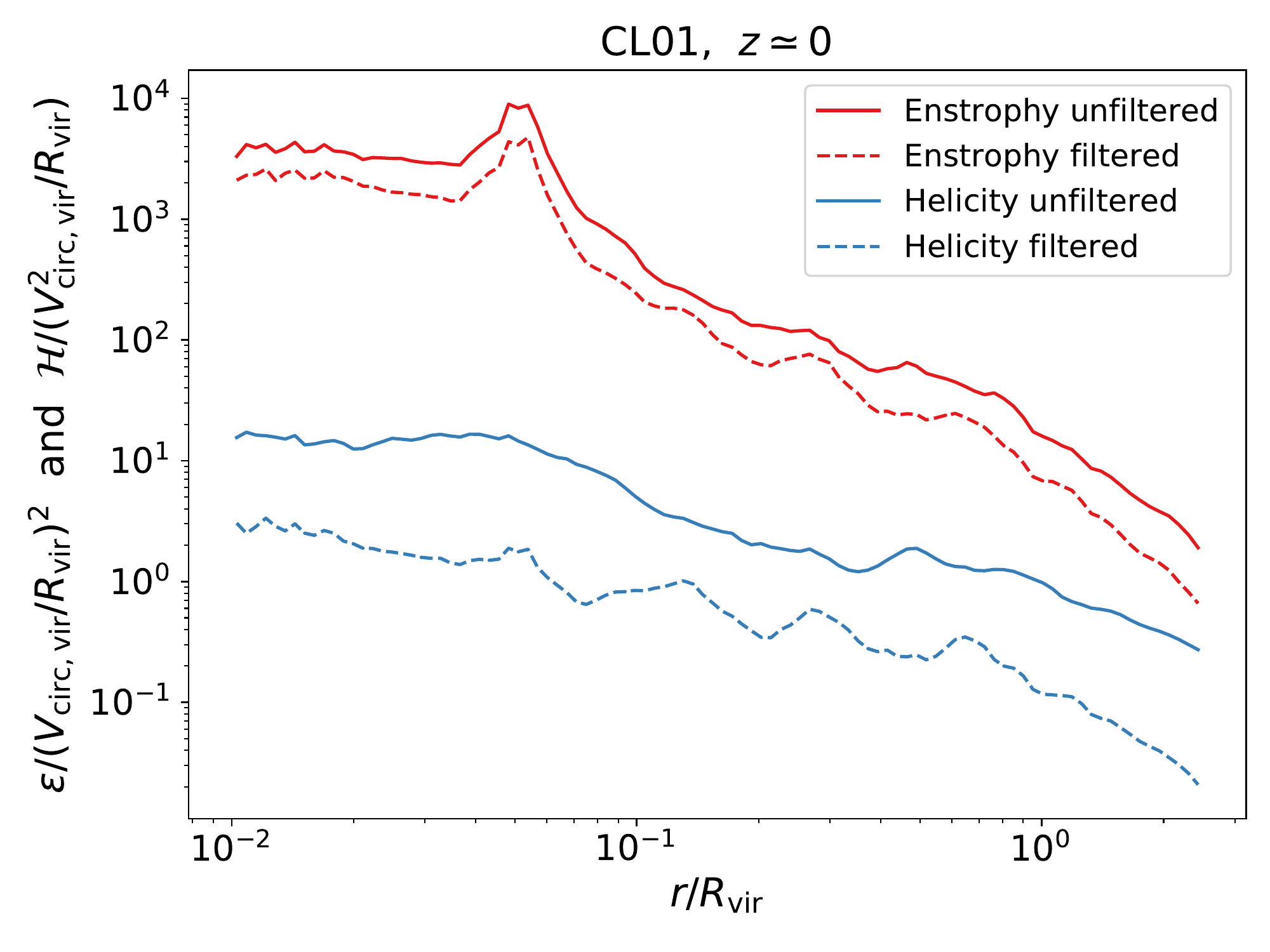}}
\caption{Radial profiles of enstrophy (red) and helicity magnitude (blue) for the cluster CL01 at $z \simeq 0$. Solid (dashed) lines correspond to the unfiltered (filtered) velocity field. The quantities are normalized in units of $R_\mathrm{vir}$ and $V_\mathrm{circ,vir} \equiv \sqrt{G M_\mathrm{tot,vir}/R_\mathrm{vir}}$.}
\label{fig:profiles_enstrophy_helicity}
\end{figure}

Complementarily, in Figure \ref{fig:profiles_enstrophy_helicity} we present radial profiles of enstrophy (red lines) and helicity magnitude (blue lines), in solid (dashed) lines for the unfiltered (filtered) velocity fields. The small-scale (or turbulent) enstrophy accounts for $\sim 50\%$ of the total enstrophy at all radii, only with a subtle decrease towards larger radii (from $\sim 60\%$ in the core to $\sim 40\%$ in the outskirts), which may well be explained by the decreased resolution in the outskirts: as the flows in these regions are typically resolved with larger cells, the multiscale filtering algorithm (Sec. \ref{s:methods.filter}) may converge to an outer-scale length, $L(\vb{x})$, a few times the cell side length, reducing the magnitude of the turbulent velocity field. However, the magnitudes of the small-scale helicity are only $(10-30)\%$ those of the unfiltered magnitude. Therefore, as suggested in Sec. \ref{s:global.vorticity_helicity}, helicity is primarily contributed by the eddies being developed within (initially) laminarly infalling flows. 

The comparison of the enstrophy and helicity magnitudes, when normalized to the cluster-wide units (i.e., we take $R_\mathrm{vir}$ and $V_\mathrm{circ,vir}$ as the length and velocity units), clearly reflects that helical motions are much more pervasive towards the cluster outskirts: the ratio between $\frac{\mathcal{H}}{V_\mathrm{circ,vir}^2 / R_\mathrm{vir}}$ and $\frac{\epsilon}{(V_\mathrm{circ,vir} / R_\mathrm{vir})^2}$ grows with radius, from below $1 \%$ in the cluster core to $\sim 30 \%$ in the off-virial region, where laminar motions due to the accretion of gas are dominant.

\subsection{Distribution of the enstrophy source terms}
\label{s:local.sources}
\begin{figure}
\centering
\mbox{\textsf{Compressive}}
\hspace*{-.5cm}
{\includegraphics[width=0.55\linewidth]{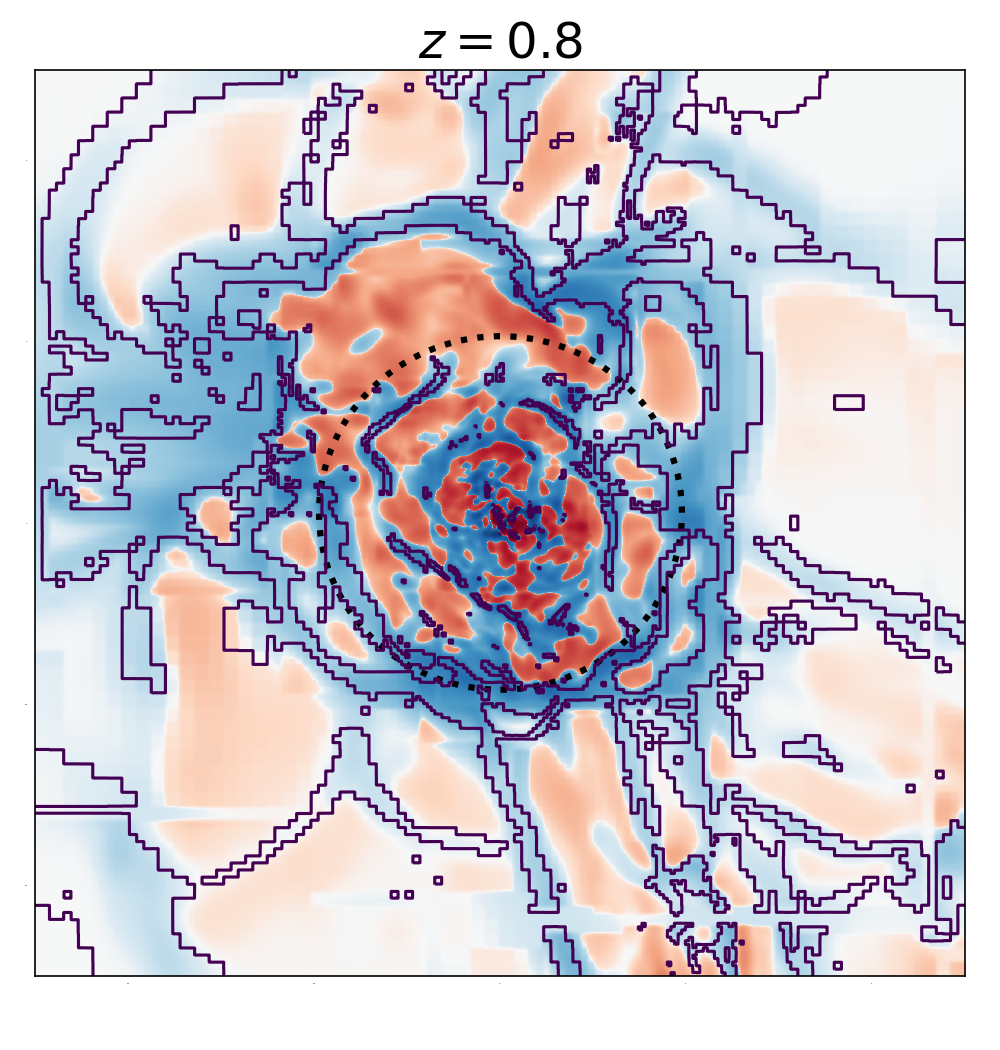}}~
{\includegraphics[width=0.55\linewidth]{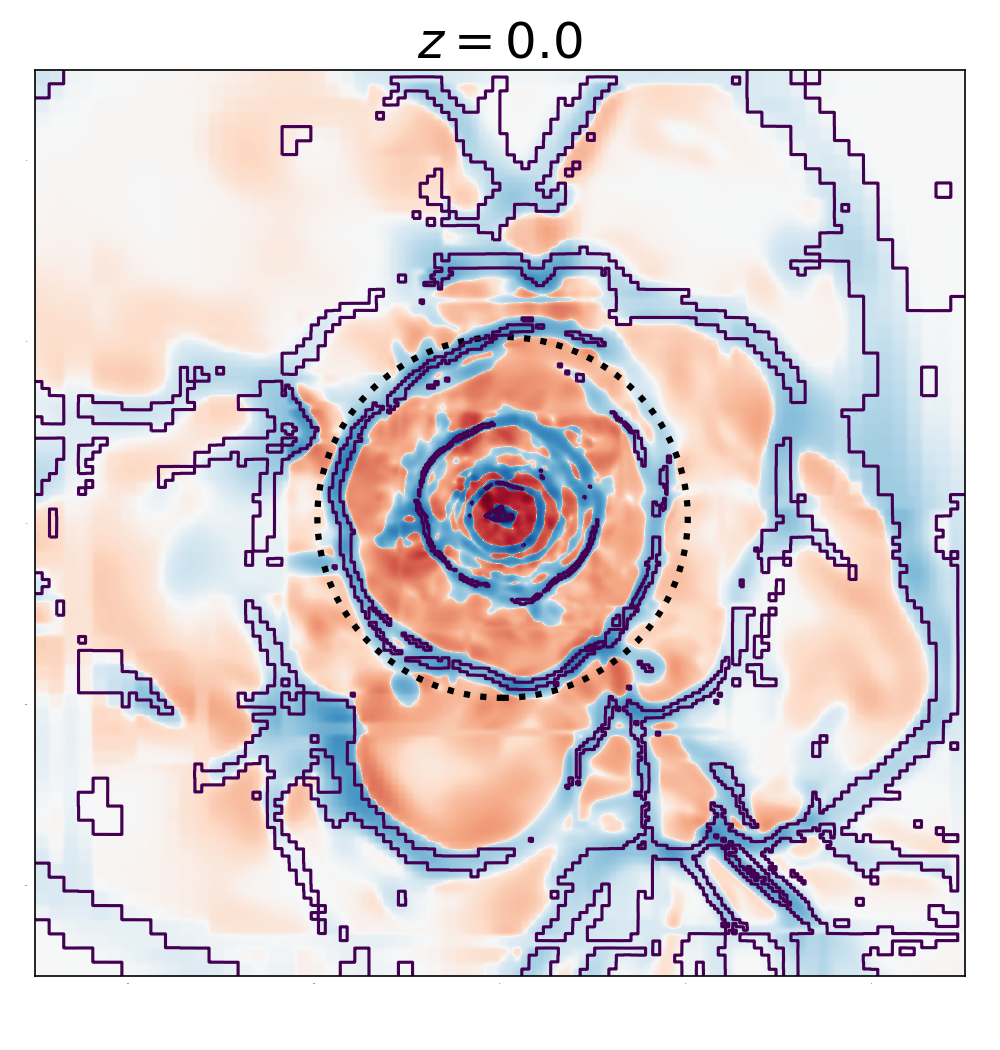}}
\mbox{\textsf{Vortex stretching}}
\hspace*{-.5cm}
{\includegraphics[width=0.55\linewidth]{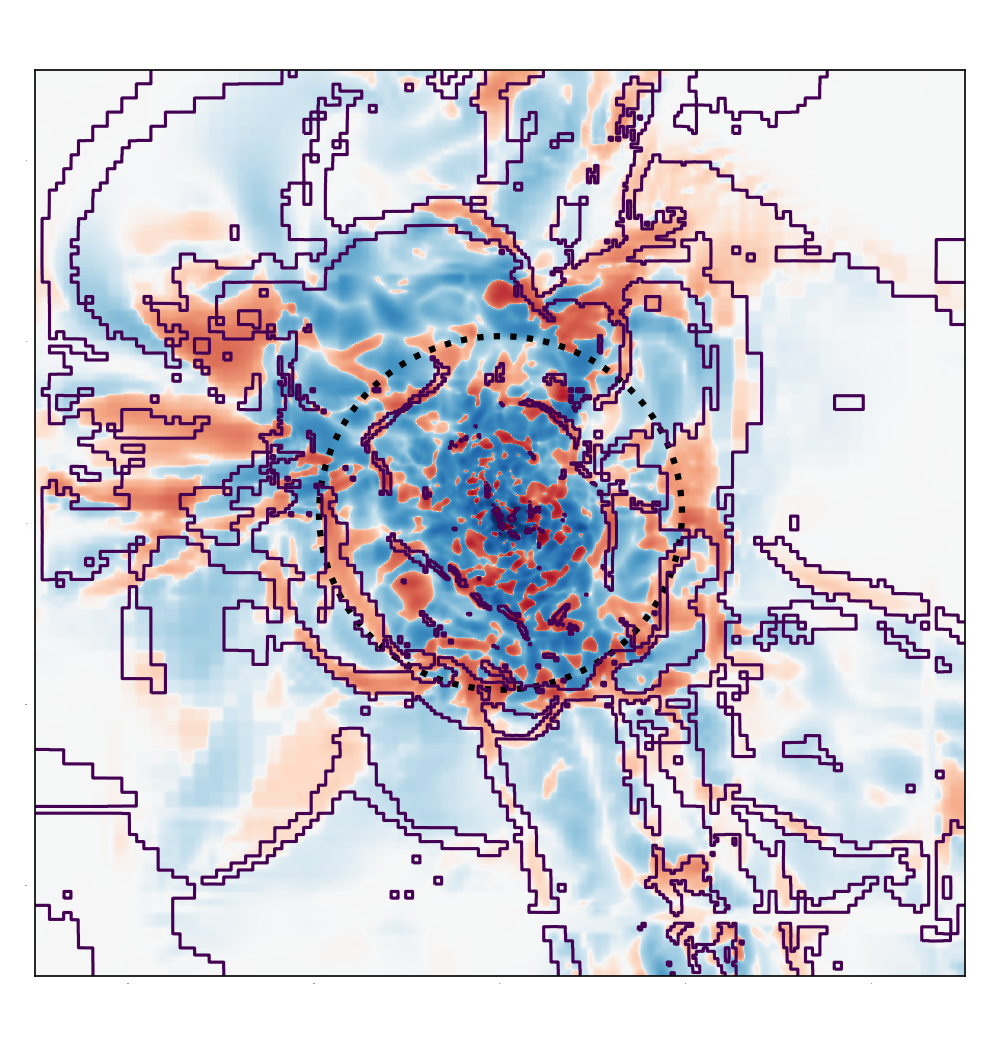}}~ 
{\includegraphics[width=0.55\linewidth]{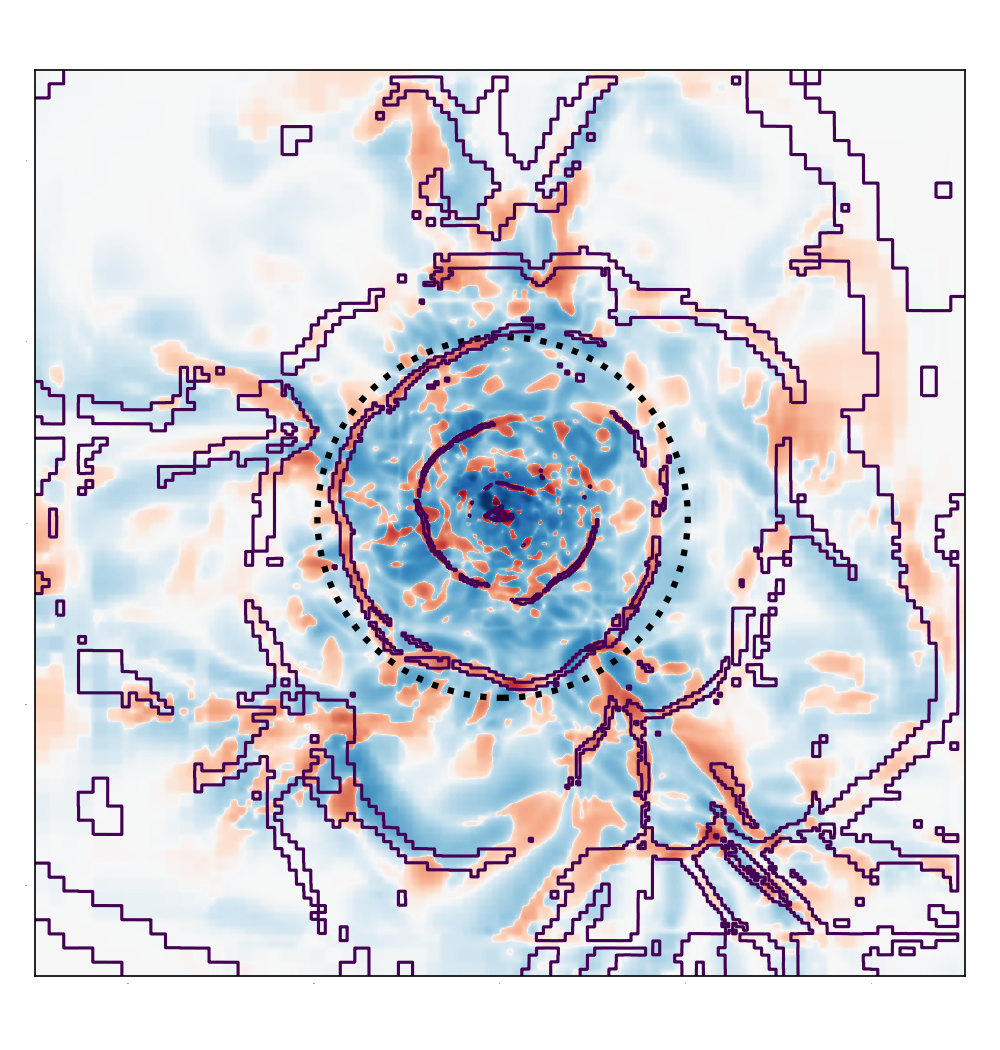}}
\mbox{\textsf{Baroclinic}}
\hspace*{-.5cm}
{\includegraphics[width=0.55\linewidth]{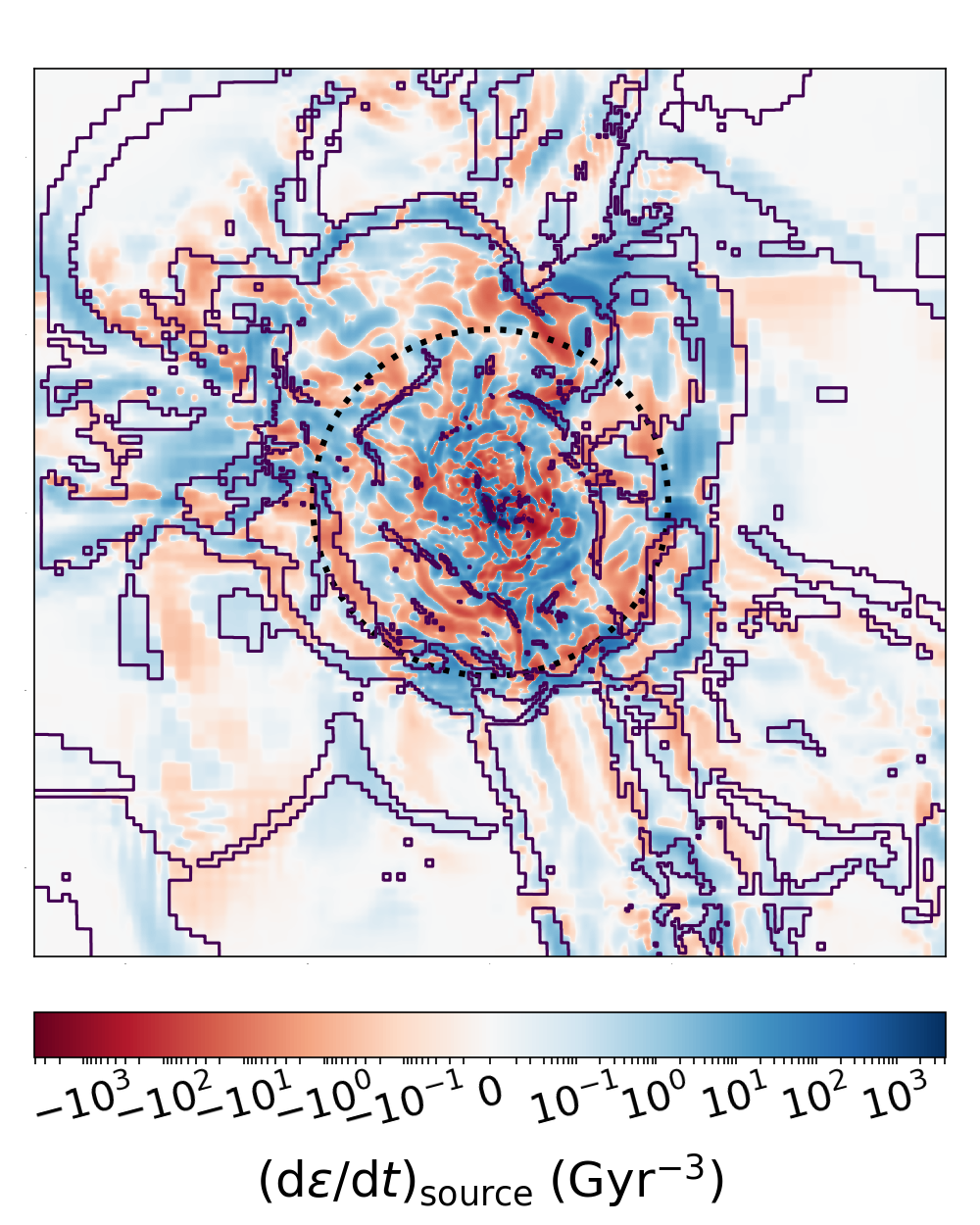}}~
{\includegraphics[width=0.55\linewidth]{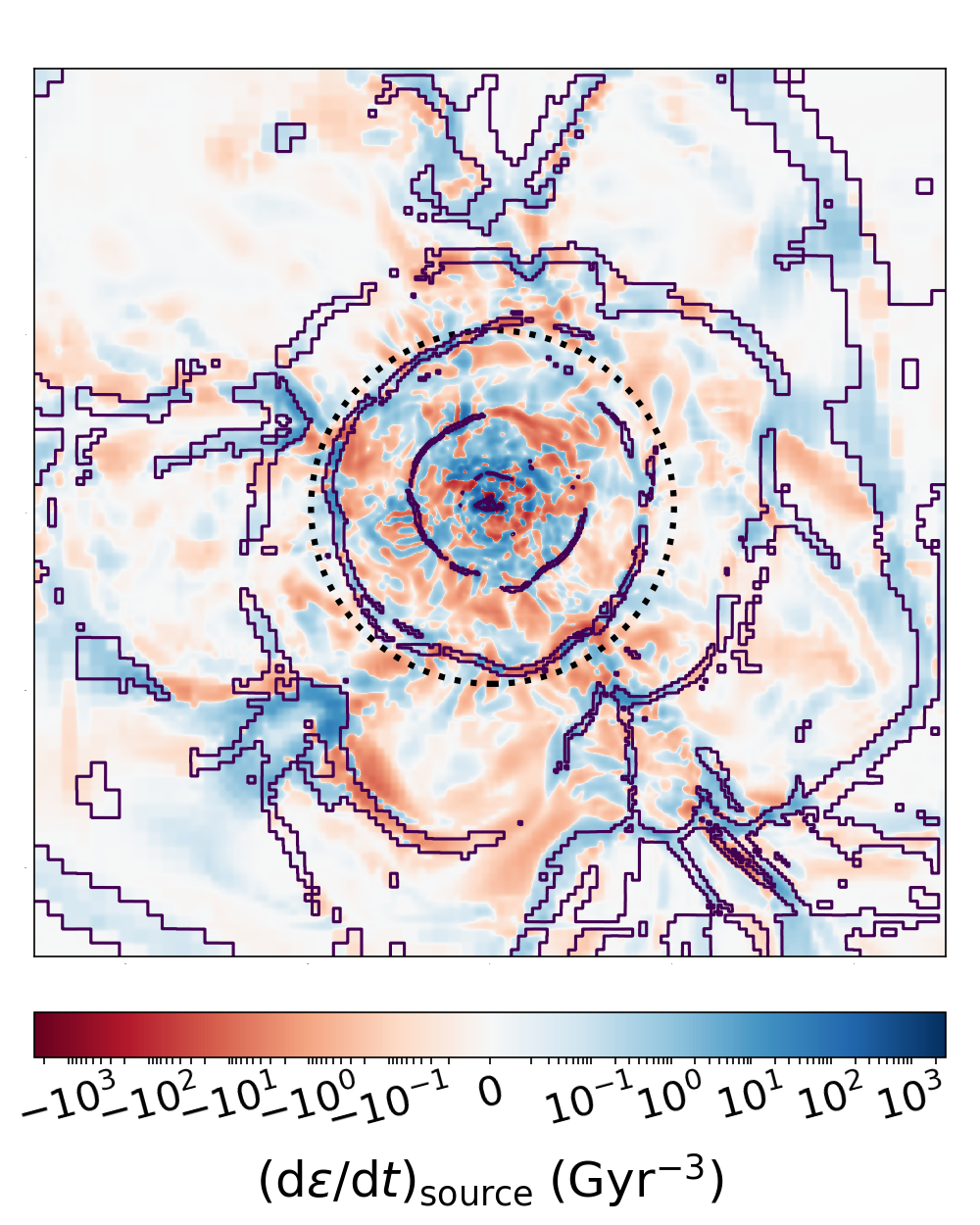}} 
\caption{Maps of the compressive (top row), vortex stretching (middle row) and baroclinic (bottom row) enstrophy source terms, for $z \simeq 0.8$ (during a major merger; left-hand panels) and for $z \simeq 0$ (smoothly accreting; right-hand panels). The slices represent the same region as those in Fig. \ref{fig:maps}. The contours indicate the shocked regions ($\mathcal{M} > 2$).}
\label{fig:maps_source terms}
\end{figure}

Complementarily to the study of the cluster-wide trends of the enstrophy source terms (Sec. \ref{s:global.vorticity_helicity.evolution}), the examination of their spatial distribution can help in distinguishing their roles in the evolution of enstrophy. Focusing on CL01, Figure \ref{fig:maps_source terms} presents the spatial distribution of the compressive (top panels), vortex stretching (middle panels) and baroclinic (bottom panels) enstrophy source terms for the same comoving region as in Fig. \ref{fig:maps}. We have focused on the snapshots at $z \simeq 0.8$ (left-hand column) and $z \simeq 0$ (right-hand column). In the maps, we have overplotted Mach number $\mathcal{M} \geq 2$ contours.

The compressive mechanism is responsible for the most part of enstrophy generation in the outskirts (beyond $R_\mathrm{vir}$), where the other source terms are smaller. This is due to the presence of strong accretion shocks (with large, negative $\nabla \cdot \vb{v}$). At more recent redshifts, the relaxed dynamical state of the cluster and its nearly spherical shape evidence an onion-like structure of enstrophy sources and sinks around the cluster. A (Lagrangian) gas element radially infalling to the cluster would experience a succession of compressions (at the accretion and merger shocks) and rarefactions, producing enstrophy increases and decreases, respectively. Since shocks are a dissipative phenomenon, this mechanism causes a net, thermodynamically irreversible generation of enstrophy (see also the Lagrangian study of \citealp{Wittor_2017}).

The main source of enstrophy in the unshocked regions, however, is the vortex stretching mechanism; while its magnitude is most typically negative inside the shock contours. This, together with the fact that this term is mostly positive for unshocked cells (see Fig. \ref{fig:phase_maps_sources} in Sec. \ref{s:global.vorticity_helicity.phase_maps}), portrays how the stretching of vortices, primarily by the accretion flows, channels the enstrophy generated in shocks to inner regions in the cluster. This fact is also supported by the correlation between the magnitude of the compressive term in shocks and that of the vortex stretching term in the immediate post-shock region. For instance, at $z \simeq 0$ the strongest compression in the external shock seems to occur in the bottom-left direction of the plot; correspondingly, the strongest vortex stretching downstream of that shock also occurs in the same direction.

Last, the baroclinic term presents complex features and variations on short spatial scales, especially towards the cluster core. Nevertheless, quite generally its value is positive in (and downstream of) the strongest shocks, in line with the behaviour observed in Sec. \ref{s:global.vorticity_helicity.phase_maps}. Its magnitude inside the virial volume of the cluster tends to be smaller than the vortex stretching and the compressive terms, especially at $z \simeq 0$.

\section{Discussion}
\label{s:discussion}

Finally, we further discuss a number of additional issues, connected to the results shown in Secs. \ref{s:global} and \ref{s:local} in relation to several models suggested in the recent literature.

In Sec. \ref{s:global.evolution}, we have focused on the velocity fluctuations over particular comoving lengths and their evolution with cosmic time. We have found the velocity fluctuations on fixed scales to be primarily determined by the presence/absence of merger events, especially major mergers, with the accretion rates determining to a fairly good extent the behaviour of these velocity fluctuations. Recently, \cite{Shi_2018} proposed a two-phase mechanism for turbulence decay after a major merger, consisting on a first phase of fast-decay, which lasts for $\sim 1 \, \mathrm{Gyr}$, and a subsequent phase of secular decay.  A similar effect is seen in our velocity fluctuation evolution graph (Fig. \ref{fig:evolution_velocity_fluctuations}). Their conclusions, as well as ours, are based on the study of a single galaxy cluster, so investigating it in large samples of clusters, with different merger configurations and accretion histories would definitely help to elucidate its physical origin. Therefore, although we indeed detect a similar behaviour, we leave this research direction for future work.

By studying the evolution of enstrophy sources through cosmic time (Sec. \ref{s:global.vorticity_helicity.evolution}), we have found the primary, cluster-wide mechanisms for enstrophy increase to be peculiar compression, vortex stretching and baroclinicity, while the rest of mechanisms tend to have more modest contributions, in general terms. However, the cluster-averaged values of the source terms do not represent the full physical picture. In order to better discriminate the conditions under which each of the mechanisms operates, we have followed two paths. 

First, we have explored the phase-space distribution of the main enstrophy source terms, mentioned above, with the shock Mach number. These three source terms have clearly distinctive behaviour with $\mathcal{M}$, which already gives information about where they do take place. Naturally, the larger the Mach number is, the more efficient the compressive mechanism is in enhancing the local enstrophy. On the other hand, vortex stretching most usually removes enstrophy from strong shocks, and increases it in unshocked regions. Last, we have seen that baroclinicity most often is a positive source of enstrophy, and its magnitude increases continuously with $\mathcal{M}$. Complementarily, we have also looked at the spatial distribution of these enstrophy source terms (Sec. \ref{s:local.sources}). 

Our results seem to add evidence supporting the mechanism for generation of enstrophy (thus, solenoidal turbulence) proposed by \cite{Vazza_2017}, and further backed by the studies of \cite{Wittor_2017} using Lagrangian passive tracer particles. This mechanism, which especially applies to smooth accretion, can be summarized in two stages:

\begin{itemize}
	\item As low-density gas, with negligible amounts of enstrophy, infalls for the first time into a cluster, enstrophy is mainly generated at the outermost, accretion shocks by the compressive and the baroclinic mechanism. This stage can repeat at inner radii, if merger shocks are present.
	\item Enstrophy is then accreted with the infalling gas. The dominant mechanism in this stage is vortex stretching, which transports the enstrophy downstream of shocks, where it had been generated. 
\end{itemize}

We note that some level of baroclinicity is always necessary in the first place, since it is the only term in Eq. \ref{eq:evolution_vorticity} capable of generating vorticity from gas with $\bm{\omega} = \vb{0}$. Note that at high redshift, in the linear regime and even in the mildly non-linear regime (as long as the \citealp{Zeldovich_1970} approximation is applicable), the velocity field can be written as the gradient of a potential flow, thus being identically irrotational. As a consequence, pristine gas from low-density regions suffers a first generation of enstrophy due to baroclinicity, most significantly in the outermost shocks, closely followed by an intense amplification due to compression.

While this two-stage mechanism describes well the generation of enstrophy by smooth accretion, the picture for mergers can be slightly different. First, the gas accreted from an infalling cluster has already some degree of enstrophy. Therefore, the advection mechanism is also important during these events (indeed, the red line in the second panel of Fig. \ref{fig:evolution_enstrophy_helicity} gets enhanced during mergers). Also, in major mergers the collision of the two ICMs generates an important amount of compression.

\section{Conclusions}
\label{s:conclusions}
The intrinsically multiscale nature of turbulent phenomena poses challenging scientific and technical (numerical) problems. While a proper description of the turbulent cascade requires high-resolution simulations, a statistically meaningful analysis demands samples of (ideally) hundreds of clusters. Performing resimulations of large samples (e.g., \citealp{Cui_2018}), or dedicated simulations including specific refinement criteria (e.g., \citealp{Vazza_2017}) or phenomenological closures to model physical dissipation (e.g., \citealp{Schmidt_2016, Iapichino_2017}) is computationally expensive. In this work, we have instead explored the possibility (and the associated limitations) of studying turbulence in galaxy clusters directly extracted from cosmological simulations, neither performing resimulations nor adding phenomenological subgrid modelling of the turbulent cascade down to the dissipation scale (which introduces uncertainties due to the involved physics and the numerical scheme). With this aim, we have focused on two massive clusters from a moderate-sized cosmological simulation. Our main conclusions and findings are summarised below:

\begin{itemize}
	\item Turbulent motions in, and around, galaxy clusters can be adequately captured by pseudo-Lagrangian AMR descriptions, relying upon the ansatz that less dense regions experience milder dynamics, provided that the mass resolution is fine enough. This assertion is key to allow the exploration of large samples of clusters at a reasonable computational cost.
	\item The non-constancy of resolution causes several systematic effects to show up in some turbulence indicators, such as structure functions, which thereby limit the extent to which computing these global statistics can be useful in our case. These effects are due to the fact that (i) the smallest scales are only resolved at certain regions (thus introducing an `environment' bias) and (ii) different AMR levels have different dissipation scales. To overcome these limitations, we have taken a complementary look, focusing on the velocity fluctuations on some particular, fixed comoving lengths (so as to avoid the effect of the previously mentioned systematics) and studying their evolution with cosmic time.
	\item Velocity fluctuations at different scales, $\sqrt{S_2(L)}$, are primarily determined by the accretion rates, showing prominent peaks in their evolution after mergers (especially, major mergers). The injection scale, although difficult to constrain (due to its non-uniqueness) can be placed around $\lesssim R_\mathrm{vir}$, in consistence with previous findings. The peaks in $\sqrt{S_2(L)}$ at smaller scales occur with a delay roughly consistent with the cascade time-scale, $L/\sqrt{S_2(L)}$. The evolution of the velocity fluctuations after merger events provides suggestive evidence in favour of the two-phase mechanism for turbulence decay proposed by \cite{Shi_2018}, although larger statistics are required to properly investigate its origin.
	\item Solenoidal (incompressible) turbulence is the dominant component in the ICM. Using enstrophy as a proxy for solenoidal turbulence, we have studied its generation mechanisms and their evolution with cosmic time, which is in turn intimately connected to the assembly history of the ICM. The evolution of the sources and their spatial distribution provide strong evidence supporting the two-stage mechanism for enstrophy generation suggested by \cite{Vazza_2017}, which combines baroclinicity and compression at shocks, with vortex stretching downstream of them. Nevertheless, we highlight that: (i) baroclinicity is responsible for the first generation of enstrophy, and (ii) while this scenario naturally accounts for enstrophy generation during smooth accretion, the situation might be notably different during mergers.
	\item Complementarily, we have introduced helicity, $\mathcal{H} \equiv \vb{v} \cdot \bm{\omega}$, as a quantity which highlights eddies being developed within bulk motions of the gas infalling into the cluster. While this magnitude is more difficult to interpret, due to not being positive defined (contrary to enstrophy), we have shown that:
	\begin{enumerate}
		\item Helical motions are irrelevant in cluster cores, in the sense that they contribute little to the total enstrophy budget, but much more pervasive towards the outskirts, associated with the fact that gas inside $\sim R_\mathrm{vir}$ presents milder bulk flows than gas in the outskirts by recent redshifts.
		\item Helicity maps highlight the presence of cosmic filaments, suggesting that matter does not fall in straight, radial trajectories to cluster, but rather with complex geometries.
		\item When computed from the small-scale, turbulent velocity field, helicity averages to zero through the cluster volume (as opposed to enstrophy). This reflects how helicity captures the interaction of bulk and turbulent velocity fields, and is suitable to investigate accretion flows.
	\end{enumerate}
	\noindent We plan to thoroughly examine the role of helicity in cluster outskirts and filaments in future works.
	
\end{itemize}

As we have mentioned through the text, the particular details about the generation and evolution of ICM turbulence are extremely sensitive to the assembly history of the cluster, as well as to its environment. Thus, one of the main shortcomings of this work is the lack of statistics. In a forthcoming work, we will extend these analyses to a large sample of massive clusters, in order to extract statistically robust conclusions about the scaling relations of turbulence-related quantities with cluster mass, the differences between varying dynamical states, etc. A thorough and deep understanding of the description of turbulent phenomena in galaxy clusters is key to explain some observable effects, such as surface brightness fluctuations, acceleration of cosmic ray particles, amplification of magnetic fields, or non-thermal pressure leading to mass bias, etc., some of which can in turn be crucial for the usage of clusters for precision cosmology.


\section*{Acknowledgements}
{We thank the anonymous referee for his/her constructive comments, which have helped to improve the quality of this manuscript. This work has been supported by the Spanish Ministerio de Ciencia e Innovaci\'on (MICINN, grant PID2019-107427GB-C33) and by the Generalitat Valenciana (grant PROMETEO/2019/071). Simulations have been carried out using the supercomputer Llu\'is Vives at the Servei d'Inform\`atica of the Universitat de Val\`encia.}. This research has made use of the following open-source packages: \textsc{NumPy} \citep{Numpy}, \textsc{SciPy} \citep{Scipy}, \textsc{matplotlib} \citep{Matplotlib} and \textsc{yt} \citep{yt}. 

\section*{Data availability}
The data underlying this article will be shared on reasonable request to the corresponding author.

\bibliographystyle{mnras}
\bibliography{mnras_turbulence}

\appendix

\section{Volume averages. Resolution dependence and convergence test}
\label{s:appendix.resolutionconvergence}

Vorticity is not a scale-invariant magnitude, neither in the classical Kolmogorov theory ($\omega_L \propto L^{-2/3}$) nor in actual simulations of ICM turbulence (see, e.g., \citealp{Vazza_2017}). Thus, in order for any results obtained in either simulations or observations to be physically meaningful, it is of utmost importance to make sure that cluster-integrated vorticity (or any similar magnitude, such as enstrophy or helicity) is defined in such a way that it does not depend strongly on numerical resolution. 

In particular, in this study we have considered the volume-averaged enstrophies and helicities. Note, however, that our refinement scheme consistently samples with increased resolution the overdense regions (where dynamics vary on shorter spatial scales). Thus, in this case\footnote{This discussion could in principle be applied to SPH data, as well.}, the volume weight of these quantities naturally privileges the denser regions, since only in these regions smaller eddies will have been able to develop. Note therefore that performing a mass-weight on this `mass-sampled' data would effectively correspond to performing a double-weight, and should not be a converging quantity as the resolution gets increased.

\begin{figure}
\centering
{\includegraphics[width=\linewidth]{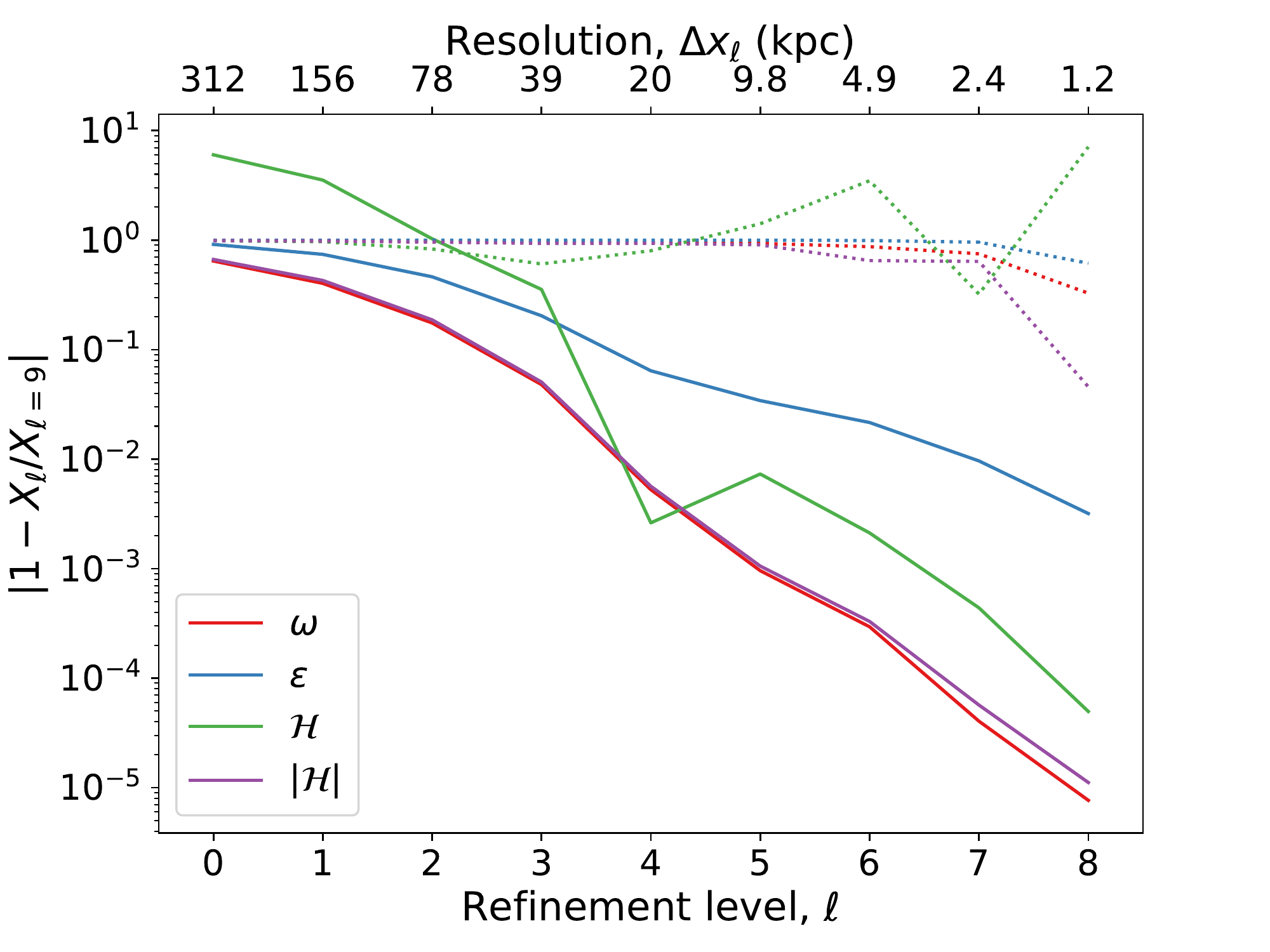}}
\caption{Resolution dependence of several cluster-averaged quantities, defined as a volume-weighted average (solid lines) or a mass-weighted average (dotted lines). The vertical axis represents the fractional variation of the quantity computed using data up to the $\ell$-th refinement level with respect to the quantity computed using the maximum resolution data at each point.}
\label{fig:integrated_resolutiondep}
\end{figure}

In order to check this is the case, we present in Figure \ref{fig:integrated_resolutiondep} the resolution dependence of several quantities (namely vorticity, enstrophy, helicity and helicity magnitude) integrated in volume and in mass, i.e. $\frac{1}{V} \int_\mathcal{V} X \dd V$ and $\frac{1}{M} \int_\mathcal{V} X \dd M$, respectively. Solid lines show the fractional difference in each quantity, when performing the volume integral up to a given refinement level, with respect to its value at the maximum available resolution ($\ell = 9$). Dotted lines present the same quantity, for the mass-weighted averages. While the mass-averages do not converge when increasing the resolution (i.e., the `errors', $|1 - X_\ell / X_{\ell = 9}|$ are always of order unity), all volume-averaged quantities do converge rapidly. Therefore, their values do not depend strongly on the resolution of the simulation and are the physically meaningful quantities in this scope.

We find necessary to emphasize that this consideration is valid as long as we work with mass-sampled simulation data. Naturally, different refinement strategies (e.g., \citealp{Vazza_2012}, who add a refinement criterion based on local vorticity) or fixed grids could rescue small-scale eddies in low-density regions, and thus enhance the volume-averaged vorticity. The only point of the reasoning above is thus to guarantee that, while we work with mass-sampled data (such as standard AMR or SPH), our results do not depend strongly on the resolution of a specific object and, thus, it is possible to compare different objects from the same or from different simulations.

\label{lastpage}
\end{document}